\documentclass{fac}
\usepackage{graphicx}
\usepackage{amssymb}
\usepackage{amsfonts}
\usepackage{amsmath}
\usepackage{dsfont}
\usepackage{lmerge}
\ifprodtf \else \usepackage{latexsym}\fi

\newcommand\black{\ensuremath{\blacktriangleright}}
\newcommand\white{\ensuremath{\vartriangleright}}

\newif\ifamsfontsloaded
\ifprodtf
  \newcommand\whbl{\white\kern-.1em--\kern-.1em\black}
  \newcommand\blwh{\black\kern-.1em--\kern-.1em\white}
  \newcommand\blbl{\black\kern-.1em--\kern-.1em\black}
  \newcommand\whwh{\white\kern-.1em--\kern-.1em\white}
  \amsfontsloadedtrue
\else
  \checkfont{msam10}
  \iffontfound
    \IfFileExists{amssymb.sty}
      {\usepackage{amssymb}\amsfontsloadedtrue
       \newcommand\whbl{\white\kern-.125em--\kern-.125em\black}%
       \newcommand\blwh{\black\kern-.125em--\kern-.125em\white}%
       \newcommand\blbl{\black\kern-.125em--\kern-.125em\black}%
       \newcommand\whwh{\white\kern-.125em--\kern-.125em\white}}
      {}
  \fi
\fi

\title[An Axiomatization for Quantum Processes to Unifying Quantum and Classical Computing]
      {An Axiomatization for Quantum Processes to Unifying Quantum and Classical Computing}

\author[Yong Wang]
    {Yong Wang\\
     School of Computer Science and Technology,\\
     Beijing University of Technology, Beijing, China\\
     }

\correspond{Yong Wang, Pingleyuan 100, Chaoyang District, Beijing, China.
            e-mail: wangy@bjut.edu.cn}

\pubyear{2011}
\pagerange{\pageref{firstpage}--\pageref{lastpage}}

\begin{document}
\label{firstpage}

\makecorrespond

\maketitle

\begin{abstract}
We establish an axiomatization for quantum processes, which is a quantum generalization of process algebra ACP (Algebra of Communicating Processes). We use the framework of a quantum process configuration $\langle p, \varrho\rangle$, but we treat it as two relative independent part: the structural part $p$ and the quantum part $\varrho$, because the establishment of a sound and complete theory is dependent on the structural properties of the structural part $p$. We let the quantum part $\varrho$ be the outcomes of execution of $p$ to examine and observe the function of the basic theory of quantum mechanics. We establish not only a strong bisimularity for quantum processes, but also a weak bisimularity to model the silent step and abstract internal computations in quantum processes. The relationship between quantum bisimularity and classical bisimularity is established, which makes an axiomatization of quantum processes possible. An axiomatization for quantum processes called qACP is designed, which involves not only quantum information, but also classical information and unifies quantum computing and classical computing. qACP can be used easily and widely for verification of most quantum communication protocols.
\end{abstract}

\begin{keywords}
Quantum Processes; Process Algebra; Algebra of Communicating Processes; Axiomatization
\end{keywords}

\section{Introduction}\label{Introduction}

The basic principles of quantum mechanics are widely adopted in computation and communication. As a relative novel computation pattern, quantum computing\cite{QCQI} brings the dawn of solving the so-called NP problem because of the strong parallel computation power of quantum computing. And also, many basic principles of quantum mechanics, such as Heisenberg uncertainty principle and quantum no-cloning theorem, provide quantum communication protocols the so-called provable security. Now, some quantum communication protocols, especially quantum key distribution protocols, have already been commercially available.

Process algebra\cite{PA} is well known in capturing traditional computation, especially parallelism and concurrence, in an interleaving pattern, such as CCS (Calculus of Concurrent Process)\cite{CCS}\cite{CCS2}, CSP (Communicating Sequential Processes)\cite{CSP} and ACP (Algebra of Communicating Processes)\cite{ACP}. To unify quantum computing and classical computing under the same process algebra framework, is attractive and has an important significance, because most quantum communication protocols involve quantum information and classical information, quantum computing and classical computing.

In this paper, we design an axiomatization called qACP for quantum processes with a quantum generalization of process algebra ACP, which unifies quantum computing and classical computing. qACP consists of not only an operational semantics based on classical structural operational semantics, but also an equational logic, by use of which, most quantum communication protocols can be verified easily.

This paper is organized as follows. In section \ref{RelatedWorks}, we introduce the related works. In section \ref{Pre}, we recall some preliminaries, including basic concepts and conclusions about basic linear algebra, basic quantum mechanics, equational logic, structural operational semantics and process algebra ACP. In section \ref{SOSQP}, we extend classical structural operational semantics to support quantum processes. The basic quantum process algebra (BQPA) is introduced in section \ref{BQPA}, Quantum Process Algebra with Parallelism (QPAP) and Algebra of Quantum Communicating Processes (AQCP) are designed in section \ref{AQCP}. To capture infinite computing in quantum processes, we discuss recursion in section \ref{Recursion}. To model silent step and abstract internal computation, silent step and abstraction operator are introduced in section \ref{Abstraction}. We unify qACP and classical ACP in section \ref{Unifying}. An example of the famous BB84 protocol\cite{BB84} is verified by use of qACP in section \ref{Verification}. qACP can be extended easily in an elegant way, which is shown in section \ref{Extensions}. Finally, we conclude this paper in section \ref{Conclusions}.

\section{Related Works}\label{RelatedWorks}

Quantum process algebra provides formal tools for modeling, analysis and verification of quantum communication protocols, which combines quantum communications and quantum computing together. \cite{CQP}\cite{CQP2} defined a language called CQP (Communicating Quantum Processes) by adding primitives for quantum measurements and transformation of quantum states to $\pi$-calculus. An operational semantics and a type system for CQP were also presented to prove that the semantics preserves typing and typing guarantees that each qubit is owned by a unique process within a system.

\cite{QPA}\cite{QPA2}\cite{PAOS}\cite{BC} defined a language called QPAlg (Quantum Process Algebra), in which, based on CCS\cite{CCS}\cite{CCS2}, primitives of unitary transformations and quantum measurements were added to CCS. An operational semantics based on probabilistic branching bisimulation was given in QPAlg.

\cite{qCCS} was introduced as a kind of algebra of pure quantum processes (no classical data involved) based on CCS. qCCS aimed at providing a suitable framework, in which the mechanism of quantum concurrent computation can be understood, and interactions and conjugation of computation and communication in quantum systems can be observed. In qCCS, quantum operations (super operators) were chosen to describe transformations of quantum states, and quantum variables and their substitutions were carefully treated. An operational semantics for qCCS based on exact (strong) bisimulation and an approximation version of bisimulation were presented for qCCS.

Based on \cite{qCCS}, several kind of bisimulations were presented for qCCS, such as probabilistic bisimulation\cite{PSQP}, a kind of weak probabilistic bisimulation\cite{BQP}, open bisimulation\cite{OBQP} and symbolic bisimulation \cite{SB}\cite{SBQP}. These bisimulations provided qCSS with more semantic models. In some bisimulations, not only pure quantum data, but also classical data could be involved in qCCS.

In this paper, we propose an axiomatization of quantum processes called qACP, which is a quantum generalization of process algebra ACP. This work uses some results of the previous works, especially qCCS, in the following ways. (1) qACP still uses the concept of a quantum process configuration $\langle p,\varrho \rangle$ \cite{PSQP} \cite{QPA} \cite{QPA2} \cite{CQP} \cite{CQP2} \cite{qCCS} \cite{BQP} \cite{PSQP} \cite{SBQP}, which is usually consisted of a process term $p$ and state information $\varrho$ of all (public) quantum information variables. (2) Like qCCS, quantum operations are chosen to describe transformations of quantum states, and behave as the atomic actions of a pure quantum process. Quantum measurements are treated as quantum operations, so probabilistic bisimulations are avoided.

There are several innovations in this paper, we enumerate them as follows. (1) A weak bisimularity (quantum branching bisimulation equivalence) is established for quantum processes. This weak bisimularity is in a non-probabilistic way that follows \cite{qCCS} and can be used to model silent step and abstract internal actions. (2) We still use the framework of a quantum process configuration $\langle p, \varrho\rangle$, but we treat it as two relative independent part: the structural part $p$ and the quantum part $\varrho$, because the establishment of a sound and complete theory is dependent on the structural properties of the structural part $p$. We let the quantum part $\varrho$ be the outcomes of execution of $p$ to examine and observe the function of the basic theory of quantum mechanics. We establish the relationship between quantum bisimularity and classical bisimularity, including strong bisimularity and weak bisimularity, which makes an axiomatization of quantum processes possible. (3) We establish a series of axiomatizations of quantum process algebra, including BQPA (Basic Quantum Process Algebra), QPAP (Quantum Process Algebra with Parallelism), AQCP (Algebra of Quantum Communicating Processes), AQCP with guarded linear recursion, and $\textrm{AQCP}_{\tau}$ with guarded linear recursion. Though these axiomatizations are based on classical axiomatizations of ACP which is based on the structural analysis the process $p$, they are not trivial and ordinary, because it is also necessary to examine if the outcomes $\varrho$ of execution of $p$ obey the basic quantum mechanics theory. For example, the associativity law of sequential composition $\cdot$, $(x\cdot y)\cdot z = x\cdot (y\cdot z)$ ($x,y,z$ range over the collection of process terms), is based on the associativity of quantum operations. And the behaviors of the silent step $\tau$ in quantum processes and that in classical processes are different under the framework of quantum process configuration $\langle p,\varrho\rangle$. (4) In this paper, qACP and classical ACP are unified under the framework of quantum process configuration $\langle p,\varrho\rangle$. This unifying means that quantum information and classical information can be mixed in qACP and quantum computing and classical computing are unified in qACP. Thus, qACP can be used widely for verification of quantum communication protocols, which involve not only quantum information, but also classical information. (5) As a result of axiomatization, qACP has not only an operational semantics, but also an equation logic which makes the qACP can be used easily. qACP also inherits the modularity of ACP, and can be extended in an elegant way.

\section{Preliminaries}\label{Pre}

For convenience of the reader, we introduce some basic concepts about basic linear algebra, basic quantum mechanics (Please refer to \cite{QCQI} for details), equational logic, structural operational semantics and process algebra ACP (Please refer to \cite{SOS} and \cite{ACP} for more details).

\subsection{Basic Linear Algebra}\label{BLA}

\textbf{Definition \ref{BLA}.1 (Hilbert space)}. An isolated physical system is associated with a Hilbert space, which is called the state space of the system. A finite-dimensional Hilbert space is a complex vector space $\mathcal{H}$ together with an inner product, which is a mapping $\langle\cdot|\cdot\rangle:\mathcal{H}\times\mathcal{H}\rightarrow \mathbf{C}$ satisfying: (1)$\langle \varphi|\varphi\rangle\geq 0$ with equality if and only if $|\varphi\rangle=0$; (2)$\langle \varphi|\psi\rangle=\langle \psi|\varphi\rangle^*$; (3) $\langle \varphi|\lambda_1\psi_1+\lambda_2\psi_2\rangle=\lambda_1\langle\varphi|\psi_1\rangle+\lambda_2\langle\varphi|\psi_2\rangle$,
 where $\mathbf{C}$ is the set of complex numbers, and $\lambda^*$ denotes the conjugate of $\lambda$ ($\lambda\in\mathbf{C}$).

\textbf{Definition \ref{BLA}.2 (Orthonormal basis)}. For any vector $|\psi\rangle$ in $\mathcal{H}$, the length $||\psi||=\sqrt{\langle\psi|\psi\rangle}$. A vector $|\psi\rangle$ with $||\psi||=1$ is called a unit vector in its state space. An orthonormal basis of a Hilbert space $\mathcal{H}$ is a basis $\{|i\rangle\}$ with

$$\langle i|j\rangle=
\begin{cases}
1& \text{if i=j,}\\
0& \text{otherwise.}
\end{cases}$$

\textbf{Definition \ref{BLA}.3 (Trace of a linear operator)}. The trace of a linear operator $A$ on $\mathcal{H}$ is defined as

$$tr(A)=\sum_{i}\langle i|A|i\rangle.$$

\textbf{Definition \ref{BLA}.4 (Tensor products)}. The state space of a composite system is the tensor product of the state space of its components. Let $\mathcal{H}_1$ and $\mathcal{H}_2$ be two Hilbert spaces, then their tensor product $\mathcal{H}_1\otimes\mathcal{H}_2$ consists of linear vectors $|\psi_1\psi_2\rangle=|\psi_1\rangle\otimes|\psi_2\rangle$, where $\psi_1\in\mathcal{H}_1$ and $\psi_2\in\mathcal{H}_2$.

For two linear operator $A_1$ on Hilbert space $\mathcal{H}_1$, $A_2$ on Hilbert space $\mathcal{H}_2$, $A_1\otimes A_2$ is defined as

$$(A_1\otimes A_2)|\psi_1\psi_2\rangle=A_1|\psi_1\rangle\otimes A_2|\psi_2\rangle$$

where $|\psi_1\rangle\in\mathcal{H}_1$ and $|\psi_2\rangle\in\mathcal{H}_2$.

Let $|\varphi\rangle=\sum_{i}\alpha_i|\varphi_{1i}\varphi_{2i}\rangle\in\mathcal{H}_1\otimes\mathcal{H}_2$ and $|\psi\rangle=\sum_{j}\beta_i|\psi_{1j}\psi_{2j}\rangle\in\mathcal{H}_1\otimes\mathcal{H}_2$. Then the inner product of $|\varphi\rangle$ and $|\psi\rangle$ is defined as follows.

$$\langle\varphi|\psi\rangle=\sum_{i,j}\alpha_i^*\beta_j\langle\varphi_{1i}|\psi_{1j}\rangle\langle\varphi_{2i}|\psi_{2j}\rangle.$$

\subsection{Basic Quantum Mechanics}\label{BQM}

\textbf{Definition \ref{BQM}.1 (Density operator)}. A mixed state of quantum system is represented by a density operator. A density operator in $\mathcal{H}$ is a linear operator $\varrho$ satisfying:(1) $\varrho$ is positive, that is, $\langle \psi|\varrho|\psi\rangle\geq 0$ for all $|\psi\rangle$; (2) $tr(\varrho)=1$. Let $\mathcal{D}(\mathcal{H})$ denote the set of all positive operators on $\mathcal{H}$.

\textbf{Definition \ref{BQM}.2 (Unitary operator)}. The evolution of a closed quantum system is described by a unitary operator on its state space. A unitary operator is a linear operator $U$ on a Hilbert space $\mathcal{H}$ with $U^{\dagger}U=\mathcal{I}_{\mathcal{H}}$, where $\mathcal{I}_{\mathcal{H}}$ is the identity operator on $\mathcal{H}$ and $U^{\dagger}$ is the adjoint of $U$.

\textbf{Definition \ref{BQM}.3 (Quantum measurement)}. A quantum measurement consists of a collection of measurement operators $\{M_m\}$, where $m$ is the measurement outcomes and satisfies

$$\sum_{m}M_{m}^{\dagger}M_m=\mathcal{I}_{\mathcal{H}}.$$

\textbf{Definition \ref{BQM}.4 (Quantum operation (super operator))}. The evolution of an open quantum system can be described by a quantum operation. A quantum operation on a Hilbert space $\mathcal{H}$ is a linear operator $\mathcal{S}$ from the space of linear operators on $\mathcal{H}$ into itself satisfying: (1) $tr[\mathcal{S}(\varrho)]\leq tr(\varrho)$ for each $\varrho\in\mathcal{D}(\mathcal{H})$; (2) for any extra Hilbert space $\mathcal{H}_R$, $(\mathcal{I}_R\otimes\mathcal{S}(A))$ is positive if $A$ is a positive operator on $\mathcal{H}_R\otimes \mathcal{H}$, where $\mathcal{I}_R$ is the identity operator in $\mathcal{H}_R$. If $tr[\mathcal{S}(\varrho)]= tr(\varrho)$ for all $\varrho\in\mathcal{D}(\mathcal{H})$, then $\mathcal{S}$ is said to be trace-preserving.

\textbf{Definition \ref{BQM}.5 (Relation between quantum operation and unitary operator)}. Let $U$ be a unitary operator on the Hilbert space $\mathcal{H}$, and $\mathcal{S}(\varrho)=U\varrho U^{\dagger}$ for any $\varrho\in\mathcal{D}(\mathcal{H})$. Then $\mathcal{S}$ is a trace-preserving quantum operation.

\textbf{Definition \ref{BQM}.6 (Relation between quantum operation and measurement operator)}. Let $\{M_m\}$ be a quantum measurement on the Hilbert space $\mathcal{H}$. For each $m$, let $\mathcal{S}_m(\varrho)=M_m\varrho M_m^{\dagger}$ for any $\varrho\in\mathcal{D}(\mathcal{H})$. The $S_m$ is a quantum operation and is not necessarily trace-preserving.

\subsection{Equational Logic}\label{ELP}

We introduce some basic concepts about equational logic briefly, including signature, term, substitution, axiomatization, equality relation, model, term rewriting system, rewrite relation, normal form, termination, weak confluence and several conclusions. These concepts are coming from \cite{ACP}, and are introduced briefly as follows. About the details, please see \cite{ACP}.

\textbf{Definition \ref{ELP}.1 (Signature)}. A signature $\Sigma$ consists of a finite set of function symbols (or operators) $f,g,\cdots$, where each function symbol $f$ has an arity $ar(f)$, being its number of arguments. A function symbol $a,b,c,\cdots$ of arity \emph{zero} is called a constant, a function symbol of arity one is called unary, and a function symbol of arity two is called binary.

\textbf{Definition \ref{ELP}.2 (Term)}. Let $\Sigma$ be a signature. The set $\mathbb{T}(\Sigma)$ of (open) terms $s,t,u,\cdots$ over $\Sigma$ is defined as the least set satisfying: (1)each variable is in $\mathbb{T}(\Sigma)$; (2) if $f\in \Sigma$ and $t_1,\cdots,t_{ar(f)}\in\mathbb{T}(\Sigma)$, then $f(t_1,\cdots,t_{ar(f)}\in\mathbb{T}(\Sigma))$. A term is closed if it does not contain variables. The set of closed terms is denoted by $\mathcal{T}(\Sigma)$.

To obey the quantum no-cloning theorem of quantum information, substitution of quantum information must be carefully treated\cite{qCCS}, which is required to be an one-to-one mapping and the passing of quantum information is always by name, but not by value. Since process algebra ACP mainly concerns the algebraic properties of actions or operations\cite{PA}, but not data or information, the substitution of terms used in this paper is just the same as classical computing. Though actions or operations manipulate data or information ultimately, it is the duty of actions or operations to obey the no-cloning theorem of quantum information.

\textbf{Definition \ref{ELP}.3 (Substitution)}. Let $\Sigma$ be a signature. A substitution is a mapping $\sigma$ from variables to the set $\mathbb{T}(\Sigma)$ of open terms. A substitution extends to a mapping from open terms to open terms: the term $\sigma(t)$ is obtained by replacing occurrences of variables $x$ in t by $\sigma(x)$. A substitution $\sigma$ is closed if $\sigma(x)\in\mathcal{T}(\Sigma)$ for all variables $x$.

\textbf{Definition \ref{ELP}.4 (Axiomatization)}. An axiomatization over a signature $\Sigma$ is a finite set of equations, called axioms, of the form $s=t$ with $s,t\in\mathbb{T}(\Sigma)$.

\textbf{Definition \ref{ELP}.5 (Equality relation)}. An axiomatization over a signature $\Sigma$ induces a binary equality relation $=$ on $\mathbb{T}(\Sigma)$ as follows. (1)(Substitution) If $s=t$ is an axiom and $\sigma$ a substitution, then $\sigma(s)=\sigma(t)$. (2)(Equivalence) The relation $=$ is closed under reflexivity, symmetry, and transitivity. (3)(Context) The relation $=$ is closed under contexts: if $t=u$ and $f$ is a function symbol with $ar(f)>0$, then $f(s_1,\cdots,s_{i-1},t,s_{i+1},\cdots,s_{ar(f)})=f(s_1,\cdots,s_{i-1},u,s_{i+1},\cdots,s_{ar(f)})$.

\textbf{Definition \ref{ELP}.6 (Model)}. Assume an axiomatization $\mathcal{E}$ over a signature $\Sigma$, which induces an equality relation $=$. A model for $\mathcal{E}$ consists of a set $\mathcal{M}$ together with a mapping $\phi: \mathcal{T}(\Sigma)\rightarrow\mathcal{M}$. (1)$(\mathcal{M},\phi)$ is sound for $\mathcal{E}$ if $s=t$ implies $\phi(s)\equiv\phi(t)$ for $s,t\in\mathcal{T}(\Sigma)$; (2)$(\mathcal{M},\phi)$ is complete for $\mathcal{E}$ if $\phi(s)\equiv\phi(t)$ implies $s=t$ for $s,t\in\mathcal{T}(\Sigma)$.

\textbf{Definition \ref{ELP}.7 (Term rewriting system)}. Assume a signature $\Sigma$. A rewrite rule is an expression $s\rightarrow t$ with $s,t\in\mathbb{T}(\Sigma)$, where: (1)the left-hand side $s$ is not a single variable; (2)all variables that occur at the right-hand side $t$ also occur in the left-hand side $s$. A term rewriting system (TRS) is a finite set of rewrite rules.

\textbf{Definition \ref{ELP}.8 (Rewrite relation)}. A TRS over a signature $\Sigma$ induces a one-step rewrite relation $\rightarrow$ on $\mathbb{T}(\Sigma)$ as follows. (1)(Substitution) If $s\rightarrow t$ is a rewrite rule and $\sigma$ a substitution, then $\sigma(s)\rightarrow\sigma(t)$. (2)(Context) The relation $\rightarrow$ is closed under contexts: if $t\rightarrow u$ and f is a funtion symbol with $ar(f)>0$, then $f(s_1,\cdots,s_{i-1},t,s_{i+1},\cdots,s_{ar(f)})\rightarrow f(s_1,\cdots,s_{i-1},u,s_{i+1},\cdots,s_{ar(f)})$. The rewrite relation $\rightarrow^*$ is the reflexive transitive closure of the one-step rewrite relation $\rightarrow$: (1) if $s\rightarrow t$, then $s\rightarrow^* t$; (2) $t\rightarrow^* t$; (3) if $s\rightarrow^* t$ and $t\rightarrow^* u$, then $s\rightarrow^* u$.

\textbf{Definition \ref{ELP}.9 (Normal form)}. A term is called a normal form for a TRS if it cannot be reduced by any of the rewrite rules.

\textbf{Definition \ref{ELP}.10 (Termination)}. A TRS is terminating if it does not induce infinite reductions $t_0\rightarrow t_1\rightarrow t_2\rightarrow \cdots$.

\textbf{Definition \ref{ELP}.11 (Weak confluence)}. A TRS is weakly confluent if for each pair of one-step reductions $s\rightarrow t_1$ and $s\rightarrow t_2$, there is a term $u$ such that $t_1\rightarrow^* u$ and $t_2\rightarrow^* u$.

\textbf{Theorem \ref{ELP}.1 (Newman's lemma)}. If a TRS is terminating and weakly confluent, then it reduces each term to a unique normal form.

\textbf{Definition \ref{ELP}.12 (Commutativity and associativity)}. Assume an axiomatization $\mathcal{E}$. A binary function symbol $f$ is commutative if $\mathcal{E}$ contains an axiom $f(x,y)=f(y,x)$ and associative if $\mathcal{E}$ contains an axiom $f(f(x,y),z)=f(x,f(y,z))$.

\textbf{Definition \ref{ELP}.13 (Convergence)}. A pair of terms $s$ and $t$ is said to be convergent if there exists a term $u$ such that $s\rightarrow^* u$ and $t\rightarrow^* u$.

Axiomatizations can give rise to TRSs that are not weakly confluent, which can be remedied by Knuth-Bendix completion\cite{KB}. It determines overlaps in left hand sides of rewrite rules, and introduces extra rewrite rules to join the resulting right hand sides, witch are called critical pairs.

\textbf{Theorem \ref{ELP}.2}. A TRS is weakly confluent if and only if all its critical pairs are convergent.

\subsection{Structural Operational Semantics}\label{SOSP}

The concepts about structural operational semantics include  labelled transition system (LTS), transition system specification (TSS), transition rule and its source, source-dependent, conservative extension, fresh operator, panth format, congruence, bisimulation, etc. These concepts are coming from \cite{ACP}, and are introduced briefly as follows. About the details, please see \cite{SOS}.

We assume a non-empty set $S$ of states, a finite, non-empty set of transition labels $A$ and a finite set of predicate symbols.

\textbf{Definition \ref{SOSP}.1 (Labeled transition system)}. A transition is a triple $(s,a,s')$ with $a\in A$, or a pair (s, P) with $P$ a predicate, where $s,s'\in S$. A labeled transition system (LTS) is possibly infinite set of transitions. An LTS is finitely branching if each of its states has only finitely many outgoing transitions.

\textbf{Definition \ref{SOSP}.2 (Transition system specification)}. A transition rule $\rho$ is an expression of the form $\frac{H}{\pi}$, with $H$ a set of expressions $t\xrightarrow{a}t'$ and $tP$ with $t,t'\in \mathbb{T}(\Sigma)$, called the (positive) premises of $\rho$, and $\pi$ an expression $t\xrightarrow{a}t'$ or $tP$ with $t,t'\in \mathbb{T}(\Sigma)$, called the conclusion of $\rho$. The left-hand side of $\pi$ is called the source of $\rho$. A transition rule is closed if it does not contain any variables. A transition system specification (TSS) is a (possible infinite) set of transition rules.

\textbf{Definition \ref{SOSP}.3 (Proof)}. A proof from a TSS $T$ of a closed transition rule $\frac{H}{\pi}$ consists of an upwardly branching tree in which all upward paths are finite, where the nodes of the tree are labelled by transitions such that: (1) the root has label $\pi$; (2) if some node has label $l$, and $K$ is the set of labels of nodes directly above this node, then (a) either $K$ is the empty set and $l\in H$, (b) or $\frac{K}{l}$ is a closed substitution instance of a transition rule in $T$.

\textbf{Definition \ref{SOSP}.4 (Generated LTS)}. We define that the LTS generated by a TSS $T$ consists of the transitions $\pi$ such that $\frac{\emptyset}{\pi}$ can be proved from $T$.

\textbf{Definition \ref{SOSP}.5}. A set $N$ of expressions $t\nrightarrow^{a}$ and $t\neg P$ (where $t$ ranges over closed terms, $a$ over $A$ and $P$ over predicates) hold for a set $\mathcal{S}$ of transitions, denoted by $\mathcal{S}\vDash N$, if: (1) for each $t\nrightarrow^{a} \in N$ we have that $t\xrightarrow{a}t' \notin \mathcal{S}$ for all $t'\in\mathcal{T}(\Sigma)$; (2) for each $t\neg P\in N$ we have that $tP \notin \mathcal{S}$.

\textbf{Definition \ref{SOSP}.6 (Three-valued stable model)}. A pair $\langle\mathcal{C},\mathcal{U}\rangle$ of disjoint sets of transitions is a three-valued stable model for a TSS $T$ if it satisfies the following two requirements: (1) a transition $\pi$ is in $\mathcal{C}$ if and only if $T$ proves a closed transition rule $\frac{N}{\pi}$ where $N$ contains only negative premises and $\mathcal{C}\cup\mathcal{U}\vDash N$; (2) a transition $\pi$ is in $\mathcal{C}\cup\mathcal{U}$ if and only if $T$ proves a closed transition rule $\frac{N}{\pi}$ where $N$ contains only negative premises and $\mathcal{C}\vDash N$.

\textbf{Definition \ref{SOSP}.7 (Ordinal number)}. The ordinal numbers are defined inductively by: (1) $0$ is the smallest ordinal number; (2) each ordinal number $\alpha$ has a successor $\alpha + 1$; (3) each sequence of ordinal number $\alpha < \alpha + 1 < \alpha + 2 < \cdots$ is capped by a limit ordinal $\lambda$.

\textbf{Definition \ref{SOSP}.8 (Positive after reduction)}. A TSS is positive after reduction if its least three-valued stable model does not contain unknown transitions.

\textbf{Definition \ref{SOSP}.9 (Stratification)}. A stratification for a TSS is a weight function $\phi$ which maps transitions to ordinal numbers, such that for each transition rule $\rho$ with conclusion $\pi$ and for each closed substitution $\sigma$: (1) for positive premises $t\xrightarrow{a}t'$ and $tP$ of $\rho$, $\phi(\sigma(t)\xrightarrow{a}\sigma(t'))\leq \phi(\sigma(\pi))$ and $\phi(\sigma(t)P\leq\phi(\sigma(\pi)))$, respectively; (2) for negative premise $t\nrightarrow^{a}$ and $t\neg P$ of $\rho$, $\phi(\sigma(t)\xrightarrow{a}t')< \phi(\sigma(\pi))$ for all closed terms $t'$ and $\phi(\sigma(t)P < \phi(\sigma(\pi)))$, respectively;

\textbf{Theorem \ref{SOSP}.1}. If a TSS allows a stratification, then it is positive after reduction.

\textbf{Definition \ref{SOSP}.10 (Process graph)}. A process (graph) $p$ is an LTS in which one state $s$ is elected to be the root. If the LTS contains a transition $s\xrightarrow{a} s'$, then $p\xrightarrow{a} p'$ where $p'$ has root state $s'$. Moreover, if the LTS contains a transition $sP$, then $pP$. (1) A process $p_0$ is finite if there are only finitely many sequences $p_0\xrightarrow{a_1}p_1\xrightarrow{a_2}\cdots\xrightarrow{a_k} P_k$. (2) A process $p_0$ is regular if there are only finitely many processes $p_k$ such that $p_0\xrightarrow{a_1}p_1\xrightarrow{a_2}\cdots\xrightarrow{a_k} P_k$.

\textbf{Definition \ref{SOSP}.11 (Bisimulation)}. A bisimulation relation $\mathcal{B}$ is a binary relation on processes such that: (1) if $p\mathcal{B}q$ and $p\xrightarrow{a}p'$ then $q\xrightarrow{a}q'$ with $p'\mathcal{B}q'$; (2) if $p\mathcal{B}q$ and $q\xrightarrow{a}q'$ then $p\xrightarrow{a}p'$ with $p'\mathcal{B}q'$; (3) if $p\mathcal{B}q$ and $pP$, then $qP$; (4) if $p\mathcal{B}q$ and $qP$, then $pP$. Two processes $p$ and $q$ are bisimilar, denoted by $p\underline{\leftrightarrow} q$, if there is a bisimulation relation $\mathcal{B}$ such that $p\mathcal{B}q$.

\textbf{Definition \ref{SOSP}.12 (Congruence)}. Let $\Sigma$ be a signature. An equivalence relation $\mathcal{B}$ on $\mathcal{T}(\Sigma)$ is a congruence if for each $f\in\Sigma$, if $s_i\mathcal{B}t_i$ for $i\in\{1,\cdots,ar(f)\}$, then $f(s_1,\cdots,s_{ar(f)})\mathcal{B}f(t_1,\cdots,t_{ar(f)})$.

\textbf{Definition \ref{SOSP}.13 (Panth format)}. A transition rule $\rho$ is in panth format if it satisfies the following three restrictions: (1) for each positive premise $t\xrightarrow{a} t'$ of $\rho$, the right-hand side $t'$ is single variable; (2) the source of $\rho$ contains no more than one function symbol; (3) there are no multiple occurrences of the same variable at the right-hand sides of positive premises and in the source of $\rho$. A TSS is said to be in panth format if it consists of panth rules only.

\textbf{Theorem \ref{SOSP}.2}. If a TSS is positive after reduction and in panth format, then the bisimulation equivalence that it induces is a congruence.

\textbf{Definition \ref{SOSP}.14 (Branching bisimulation)}. A branching bisimulation relation $\mathcal{B}$ is a binary relation on the collection of processes such that: (1) if $p\mathcal{B}q$ and $p\xrightarrow{a}p'$ then either $a\equiv \tau$ and $p'\mathcal{B}q$ or there is a sequence of (zero or more) $\tau$-transitions $q\xrightarrow{\tau}\cdots\xrightarrow{\tau}q_0$ such that $p\mathcal{B}q_0$ and $q_0\xrightarrow{a}q'$ with $p'\mathcal{B}q'$; (2) if $p\mathcal{B}q$ and $q\xrightarrow{a}q'$ then either $a\equiv \tau$ and $p\mathcal{B}q'$ or there is a sequence of (zero or more) $\tau$-transitions $p\xrightarrow{\tau}\cdots\xrightarrow{\tau}p_0$ such that $p_0\mathcal{B}q$ and $p_0\xrightarrow{a}p'$ with $p'\mathcal{B}q'$; (3) if $p\mathcal{B}q$ and $pP$, then there is a sequence of (zero or more) $\tau$-transitions $q\xrightarrow{\tau}\cdots\xrightarrow{\tau}q_0$ such that $p\mathcal{B}q_0$ and $q_0P$; (4) if $p\mathcal{B}q$ and $qP$, then there is a sequence of (zero or more) $\tau$-transitions $p\xrightarrow{\tau}\cdots\xrightarrow{\tau}p_0$ such that $p_0\mathcal{B}q$ and $p_0P$. Two processes $p$ and $q$ are branching bisimilar, denoted by $p\underline{\leftrightarrow}_b q$, if there is a branching bisimulation relation $\mathcal{B}$ such that $p\mathcal{B}q$.

\textbf{Definition \ref{SOSP}.15 (Rooted branching bisimulation)}. A rooted branching bisimulation relation $\mathcal{B}$ is a binary relation on processes such that: (1) if $p\mathcal{B}q$ and $p\xrightarrow{a}p'$ then $q\xrightarrow{a}q'$ with $p'\underline{\leftrightarrow}_b q'$; (2) if $p\mathcal{B}q$ and $q\xrightarrow{a}q'$ then $p\xrightarrow{a}p'$ with $p'\underline{\leftrightarrow}_b q'$; (3) if $p\mathcal{B}q$ and $pP$, then $qP$; (4) if $p\mathcal{B}q$ and $qP$, then $pP$. Two processes $p$ and $q$ are rooted branching bisimilar, denoted by $p\underline{\leftrightarrow}_{rb} q$, if there is a rooted branching bisimulation relation $\mathcal{B}$ such that $p\mathcal{B}q$.

\textbf{Definition \ref{SOSP}.16 (Lookahead)}. A transition rule contains lookahead if a variable occurs at the left-hand side of a premise and at the right-hand side of a premise of this rule.

\textbf{Definition \ref{SOSP}.17 (Patience rule)}. A patience rule for the i-th argument of a function symbol $f$ is a panth rule of the form

$$\frac{x_i\xrightarrow{\tau}y}{f(x_1,\cdots,x_{ar(f)})\xrightarrow{\tau}f(x_1,\cdots,x_{i-1},y,x_{i+1},\cdots,x_{ar(f)})}$$.

\textbf{Definition \ref{SOSP}.18 (RBB cool format)}. A TSS $T$ is in RBB cool format if the following requirements are fulfilled. (1) $T$ consists of panth rules that do not contain lookahead. (2) Suppose a function symbol $f$ occurs at the right-hand side the conclusion of some transition rule in $T$. Let $\rho\in T$ be a non-patience rule with source $f(x_1,\cdots,x_{ar(f)})$. Then for $i\in\{1,\cdots,ar(f)\}$, $x_i$ occurs in no more than one premise of $\rho$, where this premise is of the form $x_iP$ or $x_i\xrightarrow{a}y$ with $a\neq \tau$. Moreover, if there is such a premise in $\rho$, then there is a patience rule for the i-th argument of $f$ in $T$.

\textbf{Theorem \ref{SOSP}.3}. If a TSS is positive after reduction and in RBB cool format, then the rooted branching bisimulation equivalence that it induces is a congruence.

\textbf{Definition \ref{SOSP}.19 (Conservative extension)}. Let $T_0$ and $T_1$ be TSSs over signatures $\Sigma_0$ and $\Sigma_1$, respectively. The TSS $T_0\oplus T_1$ is a conservative extension of $T_0$ if the LTSs generated by $T_0$ and $T_0\oplus T_1$ contain exactly the same transitions $t\xrightarrow{a}t'$ and $tP$ with $t\in \mathcal{T}(\Sigma_0)$.

\textbf{Definition \ref{SOSP}.20 (Source-dependency)}. The source-dependent variables in a transition rule of $\rho$ are defined inductively as follows: (1) all variables in the source of $\rho$ are source-dependent; (2) if $t\xrightarrow{a}t'$ is a premise of $\rho$ and all variables in $t$ are source-dependent, then all variables in $t'$ are source-dependent. A transition rule is source-dependent if all its variables are. A TSS is source-dependent if all its rules are.

\textbf{Definition \ref{SOSP}.21 (Freshness)}. Let $T_0$ and $T_1$ be TSSs over signatures $\Sigma_0$ and $\Sigma_1$, respectively. A term in $\mathbb{T}(T_0\oplus T_1)$ is said to be fresh if it contains a function symbol from $\Sigma_1\setminus\Sigma_0$. Similarly, a transition label or predicate symbol in $T_1$ is fresh if it does not occur in $T_0$.

\textbf{Theorem \ref{SOSP}.4}. Let $T_0$ and $T_1$ be TSSs over signatures $\Sigma_0$ and $\Sigma_1$, respectively, where $T_0$ and $T_0\oplus T_1$ are positive after reduction. Under the following conditions, $T_0\oplus T_1$ is a conservative extension of $T_0$. (1) $T_0$ is source-dependent. (2) For each $\rho\in T_1$, either the source of $\rho$ is fresh, or $\rho$ has a premise of the form $t\xrightarrow{a}t'$ or $tP$, where $t\in \mathbb{T}(\Sigma_0)$, all variables in $t$ occur in the source of $\rho$ and $t'$, $a$ or $P$ is fresh.

\subsection{Process Algebra -- ACP}\label{ACPP}

ACP\cite{ACP} is a kind of process algebra which focuses on the specification and manipulation of process terms by use of a collection of operator symbols. In ACP, there are several kind of operator symbols, such as basic operators to build finite processes (called BPA), communication operators to express concurrency (called PAP), deadlock constants and encapsulation enable us to force actions into communications (called ACP), liner recursion to capture infinite behaviors (called ACP with linear recursion), the special constant silent step and abstraction operator (called $ACP_{\tau}$ with guarded linear recursion) allows us to abstract away from internal computations.

Bisimulation or rooted branching bisimulation based structural operational semantics is used to formally provide each process term used the above operators and constants with a process graph. The axiomatization of ACP (according the above classification of ACP, the axiomatizations are $\mathcal{E}_{\textrm{BPA}}$, $\mathcal{E}_{\textrm{PAP}}$, $\mathcal{E}_{\textrm{ACP}}$, $\mathcal{E}_{\textrm{ACP}}$ + RDP (Recursive Definition Principle) + RSP (Recursive Specification Principle), $\mathcal{E}_{\textrm{ACP}_\tau}$ + RDP + RSP + CFAR (Cluster Fair Abstraction Rule) respectively) imposes an equation logic on process terms, so two process terms can be equated if and only if their process graphs are equivalent under the semantic model.

ACP can be used to formally reason about the behaviors, such as processes executed sequentially and concurrently by use of its basic operator, communication mechanism, and recursion, desired external behaviors by its abstraction mechanism, and so on.

ACP is organized by modules and can be extended with fresh operators to express more properties of the specification for system behaviors. These extensions are required both the equational logic and the structural operational semantics to be extended. Then the extension can use the whole outcomes of ACP, such as its concurrency, recursion, abstraction, etc.

\section{Structural Operational Semantics Extended to Support Quantum Processes}\label{SOSQP}

In the above section, operational semantics are described by labelled transitions among process configurations, and a process term is enough to represent a process configuration. But in quantum processes, to avoid the abuse of quantum information which may violate the no-cloning theorem, a quantum process configuration $\langle p,\varrho \rangle$ \cite{PSQP} \cite{QPA} \cite{QPA2} \cite{CQP} \cite{CQP2} \cite{qCCS} \cite{BQP} \cite{PSQP} \cite{SBQP} is usually consisted of a process term $p$ and state information $\varrho$ of all (public) quantum information variables. Though quantum information variables are not explicitly defined in qACP and are hidden behind quantum operations, more importantly, the state information $\varrho$ is the effects of execution of a series of quantum operations on involved quantum systems, the execution of a series of quantum operations should not only obey the restrictions of the structure of the process terms, but also those of quantum mechanics principles. Through the state information $\varrho$, we can check and observe the functions of quantum mechanics principles, such as quantum entanglement, quantum measurement, etc.

So, the operational semantics of quantum processes should be defined based on quantum process configuration $\langle p,\varrho\rangle$, in which $\varrho=\varsigma$ of two state information $\varrho$ and $\varsigma$ means equality under the framework of quantum information and quantum computing, that is, these two quantum processes are in the same quantum state. Several important concepts used in this paper are following. Here, we use $\alpha, \beta$ to denote quantum operations in contrast to classical actions $a,b$.

\textbf{Definition \ref{SOSQP}.1 (Quantum process configuration)}. A quantum process configuration is defined to be a pair $\langle p,\varrho\rangle$, where $p$ is a process (graph) called structural part of the configuration, and $\varrho\in\mathcal{D}(\mathcal{H})$ specifies the current state of the environment, which is called its quantum part.

\textbf{Definition \ref{SOSQP}.2 (Quantum process graph)}. A quantum process (graph) $\langle p,\varrho\rangle$ is an LTS in which one state $s$ is elected to be the root. If the LTS contains a transition $s\xrightarrow{\alpha} s'$, then $\langle p, \varrho\rangle\xrightarrow{\alpha} \langle p',\varrho'\rangle$ where $\langle p',\varrho'\rangle$ has root state $s'$. Moreover, if the LTS contains a transition $sP$, then $\langle p,\varrho\rangle P$. (1) A quantum process $\langle p_0,\varrho_0\rangle$ is finite if and only if the process $p_0$ is finite. (2) A quantum process $\langle p_0,\varrho_0\rangle$ is regular if and only if the process $p_0$ is regular.

\textbf{Definition \ref{SOSQP}.3 (Quantum transition system specification)}. A quantum process transition rule $\rho$ is an expression of the form $\frac{H}{\pi}$, with $H$ a set of expressions $\langle t,\varrho\rangle\xrightarrow{\alpha}\langle t',\varrho'\rangle$ and $\langle t,\varrho\rangle P$ with $t,t'\in \mathbb{T}(\Sigma)$ and $\varrho,\varrho'\in\mathcal{D}(\mathcal{H})$, called the (positive) premises of $\rho$, and $t\xrightarrow{\alpha}t'$, called structural part of $H$ and denoted as $H_s$. And $\pi$ an expression $\langle t,\varrho\rangle\xrightarrow{\alpha}\langle t',\varrho'\rangle$ or $\langle t,\varrho\rangle P$ with $t,t'\in \mathbb{T}(\Sigma)$ and $\varrho,\varrho'\in\mathcal{D}(\mathcal{H})$, called the conclusion of $\rho$, and $t\xrightarrow{\alpha}t'$, called structural part of $\pi$ and denoted as $\pi_s$. The left-hand side of $\pi$ is called the source of $\rho$. $\frac{H_s}{\pi_s}$ is called the structural part of $\rho$ and denoted as $\rho_s$. A quantum process transition rule $\rho$ is closed if and only its structural part $\rho_s$ is closed. A quantum transition system specification (QTSS) is a (possible infinite) set of transition rules.

\textbf{Definition \ref{SOSQP}.4 (Quantum bisimulation)}. A bisimulation relation $\mathcal{B}$ is a binary relation on quantum processes such that: (1) if $\langle p,\varrho\rangle\mathcal{B}\langle q,\varsigma\rangle$ and $\langle p,\varrho\rangle\xrightarrow{\alpha}\langle p',\varrho'\rangle$ then $\langle q,\varsigma\rangle\xrightarrow{\alpha}\langle q',\varsigma'\rangle$ with $\langle p',\varrho'\rangle\mathcal{B}\langle q',\varsigma'\rangle$; (2) if $\langle p,\varrho\rangle\mathcal{B}\langle q,\varsigma\rangle$ and $\langle q,\varsigma\rangle\xrightarrow{\alpha}\langle q',\varsigma'\rangle$ then $\langle p,\varrho\rangle\xrightarrow{\alpha}\langle p',\varrho'\rangle$ with $\langle p',\varrho'\rangle\mathcal{B}\langle q',\varsigma'\rangle$; (3) if $\langle p,\varrho\rangle\mathcal{B}\langle q,\varsigma\rangle$ and $\langle p,\varrho\rangle P$, then $\langle q,\varsigma\rangle P$; (4) if $\langle p,\varrho\rangle\mathcal{B}\langle q,\varsigma\rangle$ and $\langle q,\varsigma\rangle P$, then $\langle p,\varrho\rangle P$. Two quantum process $\langle p,\varrho\rangle$ and $\langle q,\varsigma\rangle$ are bisimilar, denoted by $\langle p,\varrho\rangle\underline{\leftrightarrow} \langle q,\varsigma\rangle$, if there is a bisimulation relation $\mathcal{B}$ such that $\langle p,\varrho\rangle\mathcal{B}\langle q,\varsigma\rangle$.

\textbf{Definition \ref{SOSQP}.5 (Relation between quantum bisimulation and classical bisimulation)}. For two quantum processes, $\langle p,\varrho\rangle\underline{\leftrightarrow} \langle q,\varsigma\rangle$ , with $\varrho=\varsigma$, if and only if $p\underline{\leftrightarrow} q$ and $\varrho'=\varsigma'$, where $\varrho$ evolves into $\varrho'$ after execution of $p$ and $\varsigma$ evolves into $\varsigma'$ after execution of $q$.

\textbf{Definition \ref{SOSQP}.6 (Quantum branching bisimulation)}. A branching bisimulation relation $\mathcal{B}$ is a binary relation on the collection of quantum processes such that: (1) if $\langle p,\varrho\rangle\mathcal{B}\langle q,\varsigma\rangle$ and $\langle p,\varrho\rangle\xrightarrow{\alpha}\langle p',\varrho'\rangle$ then either $\alpha\equiv \tau$ and $\langle p',\varrho'\rangle\mathcal{B}\langle q,\varsigma\rangle$ or there is a sequence of (zero or more) $\tau$-transitions $\langle q,\varsigma\rangle\xrightarrow{\tau}\cdots\xrightarrow{\tau}\langle q_0,\varsigma_0\rangle$ such that $\langle p,\varrho\rangle\mathcal{B}\langle q_0,\varsigma_0\rangle$ and $\langle q_0,\varsigma_0\rangle\xrightarrow{\alpha}\langle q',\varsigma'\rangle$ with $\langle p',\varrho'\rangle\mathcal{B}\langle q',\varsigma'\rangle$; (2) if $\langle p,\varrho\rangle\mathcal{B}\langle q,\varsigma\rangle$ and $\langle q,\varsigma\rangle\xrightarrow{\alpha}\langle q',\varsigma'\rangle$ then either $\alpha\equiv \tau$ and $\langle p,\varrho\rangle\mathcal{B}\langle q',\varsigma'\rangle$ or there is a sequence of (zero or more) $\tau$-transitions $\langle p,\varrho\rangle\xrightarrow{\tau}\cdots\xrightarrow{\tau}\langle p_0,\varrho_0\rangle$ such that $\langle p_0,\varrho_0\rangle\mathcal{B}\langle q,\varsigma\rangle$ and $\langle p_0,\varrho_0\rangle\xrightarrow{\alpha}\langle p',\varrho'\rangle$ with $\langle p',\varrho'\rangle\mathcal{B}\langle q',\varsigma'\rangle$; (3) if $\langle p,\varrho\rangle\mathcal{B}\langle q,\varsigma\rangle$ and $\langle p,\varrho\rangle P$, then there is a sequence of (zero or more) $\tau$-transitions $\langle q,\varsigma\rangle\xrightarrow{\tau}\cdots\xrightarrow{\tau}\langle q_0,\varsigma_0\rangle$ such that $\langle p,\varrho\rangle\mathcal{B}\langle q_0,\varsigma_0\rangle$ and $\langle q_0,\varsigma_0\rangle P$; (4) if $\langle p,\varrho\rangle\mathcal{B}\langle q,\varsigma\rangle$ and $\langle q,\varsigma\rangle P$, then there is a sequence of (zero or more) $\tau$-transitions $\langle p,\varrho\rangle\xrightarrow{\tau}\cdots\xrightarrow{\tau}\langle p_0,\varrho_0\rangle$ such that $\langle p_0,\varrho_0\rangle\mathcal{B}\langle q,\varsigma\rangle$ and $\langle p_0,\varrho_0\rangle P$. Two quantum processes $\langle p,\varrho\rangle$ and $\langle q,\varsigma\rangle$ are branching bisimilar, denoted by $\langle p,\varrho\rangle\underline{\leftrightarrow}_b \langle q,\varsigma\rangle$, if there is a branching bisimulation relation $\mathcal{B}$ such that $\langle p,\varrho\rangle\mathcal{B}\langle q,\varsigma\rangle$.

\textbf{Definition \ref{SOSQP}.7 (Relation between quantum branching bisimulation and classical branching bisimulation)}. For two quantum processes, $\langle p,\varrho\rangle\underline{\leftrightarrow}_b \langle q,\varsigma\rangle$, with $\varrho=\varsigma$, if and only if $p\underline{\leftrightarrow}_b q$ and $\varrho'=\varsigma'$, where $\varrho$ evolves into $\varrho'$ after execution of $p$ and $\varsigma$ evolves into $\varsigma'$ after execution of $q$.

\textbf{Definition \ref{SOSQP}.8 (Quantum rooted branching bisimulation)}. A rooted branching bisimulation relation $\mathcal{B}$ is a binary relation on quantum processes such that: (1) if $\langle p,\varrho\rangle\mathcal{B}\langle q,\varsigma\rangle$ and $\langle p,\varrho\rangle\xrightarrow{\alpha}\langle p',\varrho'\rangle$ then $\langle q,\varsigma\rangle\xrightarrow{\alpha}\langle q',\varsigma'\rangle$ with $\langle p',\varrho'\rangle\underline{\leftrightarrow}_b \langle q',\varsigma'\rangle$; (2) if $\langle p,\varrho\rangle\mathcal{B}\langle q,\varsigma\rangle$ and $\langle q,\varsigma\rangle\xrightarrow{\alpha}\langle q',\varsigma'\rangle$ then $\langle p,\varrho\rangle\xrightarrow{\alpha}\langle p',\varrho'\rangle$ with $\langle p',\varrho'\rangle\underline{\leftrightarrow}_b \langle q',\varsigma'\rangle$; (3) if $\langle p,\varrho\rangle\mathcal{B}\langle q,\varsigma\rangle$ and $\langle p,\varrho\rangle P$, then $\langle q,\varsigma\rangle P$; (4) if $\langle p,\varrho\rangle\mathcal{B}\langle q,\varsigma\rangle$ and $\langle q,\varsigma\rangle P$, then $\langle p,\varrho\rangle P$. Two quantum processes $\langle p,\varrho\rangle$ and $\langle q,\varsigma\rangle$ are rooted branching bisimilar, denoted by $\langle p,\varrho\rangle\underline{\leftrightarrow}_{rb} \langle q,\varsigma\rangle$, if there is a rooted branching bisimulation relation $\mathcal{B}$ such that $\langle p,\varrho\rangle\mathcal{B}\langle q,\varsigma\rangle$.

\textbf{Definition \ref{SOSQP}.9 (Relation between quantum rooted branching bisimulation and classical rooted branching bisimulation)}. For two quantum processes, $\langle p,\varrho\rangle\underline{\leftrightarrow}_{rb} \langle q,\varsigma\rangle$, with $\varrho=\varsigma$, if and only if $p\underline{\leftrightarrow}_{rb} q$ and $\varrho'=\varsigma'$, where $\varrho$ evolves into $\varrho'$ after execution of $p$ and $\varsigma$ evolves into $\varsigma'$ after execution of $q$.

\textbf{Definition \ref{SOSQP}.10 (Congruence)}. Let $\Sigma$ be a signature and $\mathcal{D}(\mathcal{H})$ be the state space of the environment. An equivalence relation $\mathcal{B}$ on $\langle t\in\mathcal{T}(\Sigma), \varrho\in\mathcal{D}(\mathcal{H})\rangle$ is a congruence, i.e., for each $f\in\Sigma$, if $\langle s_i,\varrho_i\rangle\mathcal{B}\langle t_i, \varsigma_i\rangle$ for $i\in\{1,\cdots,ar(f)\}$, then $f(\langle s_1, \varrho_1\rangle,\cdots,\langle s_{ar(f)}, \varrho_{ar(f)}\rangle)\mathcal{B}f(\langle t_1, \varsigma_1 \rangle,\cdots,\langle t_{ar(f)}, \varsigma_{ar(f)})$. An equivalence relation $\mathcal{B}$ on $\langle t\in\mathcal{T}(\Sigma), \varrho\in\mathcal{D}(\mathcal{H})\rangle$ is a congruence, if for each $f\in\Sigma$, $s_i\mathcal{B}t_i$ for $i\in\{1,\cdots,ar(f)\}$, and $f(s_1,\cdots,s_{ar(f)})\mathcal{B}f(t_1,\cdots,t_{ar(f)})$.

\textbf{Definition \ref{SOSQP}.11 (Quantum conservative extension)}. Let $T_0$ and $T_1$ be QTSSs over signature $\Sigma_0$ and $\mathcal{D}(\mathcal{H}_0)$, and $\Sigma_1$ and $\mathcal{D}(\mathcal{H}_1)$, respectively. The QTSS $T_0\oplus T_1$ is a conservative extension of $T_0$ if the LTSs generated by $T_0$ and $T_0\oplus T_1$ contain exactly the same transitions $\langle t, \varrho\rangle\xrightarrow{\alpha}\langle t', \varrho'\rangle$ and $\langle t, \varrho\rangle P$ with $t\in \mathcal{T}(\Sigma_0)$ and $\varrho\in\mathcal{D}(\mathcal{H}_0)$, and $T_0\oplus T_1=\langle \Sigma_0 \cup \Sigma_1, \mathcal{D}(\mathcal{H}_0\otimes\mathcal{H}_1)\rangle$.

\textbf{Definition \ref{SOSQP}.12 (Relation between quantum conservative extension and classical conservative extension)}. The QTSS $T_0\oplus T_1$ is a quantum conservative extension of $T_0$ with transitions $\langle t, \varrho\rangle\xrightarrow{\alpha}\langle t', \varrho'\rangle$ and $\langle t, \varrho\rangle P$, if its corresponding TSS $T'_0\oplus T'_1$ is a conservative extension of $T'_0$ with transitions $t \xrightarrow{\alpha}t'$ and $tP$.

\section{BQPA -- Basic Quantum Process Algebra}\label{BQPA}

In the following, the variables $x,x',y,y',z,z'$ range over the collection of process terms, the variables $\upsilon,\omega$ range over the set $A$ of atomic quantum operations, $\alpha,\beta\in A$, $s,s',t,t'$ are closed items, $\tau$ is the special constant silent step, $\delta$ is the special constant deadlock, and the predicate $\xrightarrow{\alpha}\surd$ represents successful termination after execution of the quantum operation $\alpha$.

BQPA includes three kind of operators: the execution of atomic quantum operation $\alpha$, the alternative composition operator $+$ and the sequential composition operator $\cdot$. Each finite process can be represented by a closed term that is built from the set $A$ of atomic quantum operations, the alternative composition operator $+$, and the sequential composition operator $\cdot$. The collection of all basic process terms is called Basic Quantum Process Algebra (BQPA), which is abbreviated to BQPA.

\subsection{Transition Rules of BQPA}

We give the transition rules under quantum transition system specification (QTSS) for BQPA as follows.

$$\frac{}{\langle\upsilon,\varrho\rangle\xrightarrow{\upsilon}\langle\surd,\upsilon(\varrho)\rangle}$$

$$\frac{\langle x,\varrho\rangle\xrightarrow{\upsilon}\langle\surd,\varrho'\rangle}{\langle x+y,\varrho\rangle\xrightarrow{\upsilon}\langle\surd,\varrho'\rangle}$$

$$\frac{\langle x,\varrho\rangle\xrightarrow{\upsilon}\langle x',\varrho'\rangle}{\langle x+y,\varrho\rangle\xrightarrow{\upsilon}\langle x',\varrho'\rangle}$$

$$\frac{\langle y,\varrho\rangle\xrightarrow{\upsilon}\langle\surd,\varrho'\rangle}{\langle x+y,\varrho\rangle\xrightarrow{\upsilon}\langle\surd,\varrho'\rangle}$$

$$\frac{\langle y,\varrho\rangle\xrightarrow{\upsilon}\langle y',\varrho'\rangle}{\langle x+y,\varrho\rangle\xrightarrow{\upsilon}\langle y',\varrho'\rangle}$$

$$\frac{\langle x,\varrho\rangle\xrightarrow{\upsilon}\langle\surd,\varrho'\rangle}{\langle x\cdot y,\varrho\rangle\xrightarrow{\upsilon}\langle y,\varrho'\rangle}$$

$$\frac{\langle x,\varrho\rangle\xrightarrow{\upsilon}\langle x',\varrho'\rangle}{\langle x\cdot y,\varrho\rangle\xrightarrow{\upsilon}\langle x'\cdot y,\varrho'\rangle}$$

where $\upsilon(\varrho)$ represents the new state of a quantum system, whose origin state is $\varrho$, after the execution of the atomic quantum operation $\upsilon$.

\begin{itemize}
  \item The first transition rule says that each atomic quantum operation $\upsilon$ can terminate successfully, and the state of the environment would be changed from $\varrho$ to $\upsilon(\varrho)$.
  \item The next four transition rules say that $s+t$ can execute alternatively, that is, it can execute either $s$ or $t$.
  \item The last two transition rules say that $s\cdot t$ can execute sequentially, that is, it executes $s$ in the first, after successful termination of $s$, then execution of $t$ follows.
\end{itemize}

\subsection{Axiomatization for BQPA}

We design an axiomatization $\mathcal{E}_{\textrm{BQPA}}$ for BQPA modulo quantum bisimulation equivalence as Table \ref{AxiomForBQPA} shows.

\begin{center}
\begin{table}
  \begin{tabular}{@{}ll@{}}
\hline No. &Axiom\\
  QA1 & $x + y = y + x$ \\
  QA2 & $(x + y) + z = x + (y + z)$ \\
  QA3 & $x + x = x$ \\
  QA4 & $(x + y)\cdot z = x\cdot z + y\cdot z$ \\
  QA5 & $(x\cdot y)\cdot z = x\cdot (y\cdot z)$\\
\end{tabular}
\caption{Axioms for BQPA}
\label{AxiomForBQPA}
\end{table}
\end{center}

Several important conclusions are following.

\textbf{Theorem \ref{BQPA}.1}. Quantum bisimulation equivalence is a congruence with respect to BQPA.

\begin{proof}
The structural part of QTSSs for BQPA are all in panth format, so bisimulation equivalence that they induce is a congruence. According to the definition of quantum bisimulation, quantum bisimulation equivalence that QTSSs for BQPA induce is also a congruence.
\end{proof}

\textbf{Theorem \ref{BQPA}.2}. $\mathcal{E}_{\textrm{BQPA}}$ is sound for BQPA modulo quantum bisimulation equivalence.

\begin{proof}
Since quantum bisimulation is both an equivalence and a congruence for BQPA, only the soundness of the first clause in the definition of the relation $=$ is needed to be checked. That is, if $s=t$ is an axiom in $\mathcal{E}_{\textrm{BQPA}}$ and $\sigma$ a closed substitution that maps the variable in $s$ and $t$ to basic quantum process terms, then we need to check that $\langle\sigma(s),\varrho\rangle\underline{\leftrightarrow}\langle\sigma(t),\varsigma\rangle$.

Since axioms in $\mathcal{E}_{\textrm{BQPA}}$ (same as $\mathcal{E}_{\textrm{BPA}}$) are sound for BPA modulo bisimulation equivalence, according to the definition of quantum bisimulation, we only need to check if $\varrho'=\varsigma'$, where $\varrho$ evolves into $\varrho'$ after execution of $\sigma(s)$ and $\varsigma$ evolves into $\varsigma'$ after execution of $\sigma(t)$. For example, the axiom QA5 is sound for BQPA modulo quantum bisimulation equivalence, based on the associativity of quantum operations, that is, $(\sigma(s)\cdot\sigma(t))\cdot\sigma(u)(\varrho)=\sigma(s)\cdot(\sigma(t)\cdot\sigma(u))(\varsigma)$ for any $\varrho=\varsigma$.
\end{proof}

\textbf{Theorem \ref{BQPA}.3}. $\mathcal{E}_{\textrm{BQPA}}$ is complete for BQPA modulo quantum bisimulation equivalence.

\begin{proof}
To prove that $\mathcal{E}_{\textrm{BQPA}}$ is complete for BQPA modulo quantum bisilumation equivalence, it means that $\langle s, \varrho\rangle\underline{\leftrightarrow} \langle t,\varsigma\rangle$ implies $s=t$.

It was already proved that $\mathcal{E}_{\textrm{BQPA}}$ (same as $\mathcal{E}_{\textrm{BPA}}$) is complete for BPA modulo bisimulation equivalence, that is, $s\underline{\leftrightarrow} t$ implies $s=t$. $\langle s, \varrho\rangle\underline{\leftrightarrow} \langle t,\varsigma\rangle$ with $\varrho=\varsigma$ means that $s\underline{\leftrightarrow} t$ with $\varrho=\varsigma$ and $\varrho'=\varsigma'$, where $\varrho$ evolves into $\varrho'$ after execution of $s$ and $\varsigma$ evolves into $\varsigma'$ after execution of $t$, according to the definition of quantum bisimulation equivalence. The completeness of $\mathcal{E}_{\textrm{BQPA}}$ for BPA modulo bisimulation equivalence determines that $\mathcal{E}_{\textrm{BQPA}}$ is complete for BQPA modulo quantum bisimulation equivalence.
\end{proof}

\section{AQCP -- Algebra of Quantum Communicating Processes}\label{AQCP}

It is well known that process algebra captures parallelism and concurrency by means of the so-called interleaving pattern in contrast to the so-called true concurrency. Quantum processes can execute in parallel and communicate with each other, since the actions used to communicate are not quantum operations, which means that after the execution of communicating actions, the quantum state maintains unchanged. We introduce a new set $C$ of atomic communicating actions. A merge operator $\parallel$ and a communication function $\gamma: C\times C\rightarrow C$ can be used to capture the parallelism and the communication.

In the following, the variables $\upsilon,\omega$ range over the set $A$ of atomic quantum operations, and the variable $\nu,\mu$ range over the set $C$ of atomic communicating actions.

The merge $\langle s\parallel t, \varrho\rangle$ can choose to execute an initial transition of process term $s$ or an initial transition of process term $t$, and change the quantum state, which is captured by the following four transition rules.

$$\frac{\langle x, \varrho\rangle\xrightarrow{\upsilon}\langle\surd,\varrho'\rangle}{\langle x\parallel y,\varrho\rangle\xrightarrow{\upsilon}\langle y, \varrho'\rangle}$$

$$\frac{\langle x, \varrho\rangle\xrightarrow{\upsilon}\langle x',\varrho'\rangle}{\langle x\parallel y,\varrho\rangle\xrightarrow{\upsilon}\langle x'\parallel y, \varrho'\rangle}$$

$$\frac{\langle y, \varrho\rangle\xrightarrow{\upsilon}\langle\surd,\varrho'\rangle}{\langle x\parallel y,\varrho\rangle\xrightarrow{\upsilon}\langle x, \varrho'\rangle}$$

$$\frac{\langle y, \varrho\rangle\xrightarrow{\upsilon}\langle y',\varrho'\rangle}{\langle x\parallel y,\varrho\rangle\xrightarrow{\upsilon}\langle x\parallel y', \varrho'\rangle}$$

And also the merge $\langle s \parallel t, \varrho\rangle$ can choose to execute a communication of initial transitions of the process term $s$ and $t$, and does not change the quantum state, which is expressed by the following four transition rules.

$$\frac{\langle x, \varrho\rangle\xrightarrow{\nu}\langle\surd,\varrho\rangle\quad \langle y, \varrho\rangle\xrightarrow{\mu}\langle\surd,\varrho\rangle}{\langle x\parallel y,\varrho\rangle\xrightarrow{\gamma(\nu,\mu)}\langle \surd, \varrho\rangle}$$

$$\frac{\langle x, \varrho\rangle\xrightarrow{\nu}\langle\surd,\varrho\rangle\quad \langle y, \varrho\rangle\xrightarrow{\mu}\langle y',\varrho\rangle}{\langle x\parallel y,\varrho\rangle\xrightarrow{\gamma(\nu,\mu)}\langle y', \varrho\rangle}$$

$$\frac{\langle x, \varrho\rangle\xrightarrow{\nu}\langle x',\varrho\rangle\quad \langle y, \varrho\rangle\xrightarrow{\mu}\langle\surd,\varrho\rangle}{\langle x\parallel y,\varrho\rangle\xrightarrow{\gamma(\nu,\mu)}\langle x', \varrho\rangle}$$

$$\frac{\langle x, \varrho\rangle\xrightarrow{\nu}\langle x',\varrho\rangle\quad \langle y, \varrho\rangle\xrightarrow{\mu}\langle y',\varrho\rangle}{\langle x\parallel y,\varrho\rangle\xrightarrow{\gamma(\nu,\mu)}\langle x'\parallel y', \varrho\rangle}$$

\subsection{Left Merge and Communication Merge}

Since there does not exist a sound and complete finite axiomatization for BPA extended with the merge, modulo bisimulation equivalence, it is can be proved that there does not exist a sound and complete axiomatization for BQPA extended with the merge modulo quantum bisimulation equivalence either. This can be overcome by defining two extra operator that are called left merge $\leftmerge$ and communication merge $\mid$. We call BQPA extended with the merge operator $\parallel$, the left merge operator $\leftmerge$ and the communication merge operator $\mid$ as Quantum Process Algebra with Parallelism, which is abbreviated to QPAP.

\subsubsection{Transition Rules of QPAP}

The left merge $\langle s\leftmerge t, \varrho\rangle$ takes its initial transition from the process term $s$ and changes the quantum state, and then behaves as the merge $\parallel$, which is expressed by the following two transition rules.

$$\frac{\langle x, \varrho\rangle\xrightarrow{\upsilon}\langle\surd,\varrho'\rangle}{\langle x\leftmerge y,\varrho\rangle\xrightarrow{\upsilon}\langle y, \varrho'\rangle}$$

$$\frac{\langle x, \varrho\rangle\xrightarrow{\upsilon}\langle x',\varrho'\rangle}{\langle x\leftmerge y,\varrho\rangle\xrightarrow{\upsilon}\langle x'\parallel y, \varrho'\rangle}$$

The communication merge $\langle s\mid t, \varrho\rangle$ executes as initial transition a communication between initial transition of the process term $s$ and $t$, and does not change the quantum state, and then behaves as the merge operator $\parallel$, which is captured by the following four transition rules.

$$\frac{\langle x, \varrho\rangle\xrightarrow{\nu}\langle\surd,\varrho\rangle\quad \langle y, \varrho\rangle\xrightarrow{\mu}\langle\surd,\varrho\rangle}{\langle x\mid y,\varrho\rangle\xrightarrow{\gamma(\nu,\mu)}\langle \surd, \varrho\rangle}$$

$$\frac{\langle x, \varrho\rangle\xrightarrow{\nu}\langle\surd,\varrho\rangle\quad \langle y, \varrho\rangle\xrightarrow{\mu}\langle y',\varrho\rangle}{\langle x\mid y,\varrho\rangle\xrightarrow{\gamma(\nu,\mu)}\langle y', \varrho\rangle}$$

$$\frac{\langle x, \varrho\rangle\xrightarrow{\nu}\langle x',\varrho\rangle\quad \langle y, \varrho\rangle\xrightarrow{\mu}\langle\surd,\varrho\rangle}{\langle x\mid y,\varrho\rangle\xrightarrow{\gamma(\nu,\mu)}\langle x', \varrho\rangle}$$

$$\frac{\langle x, \varrho\rangle\xrightarrow{\nu}\langle x',\varrho\rangle\quad \langle y, \varrho\rangle\xrightarrow{\mu}\langle y',\varrho\rangle}{\langle x\mid y,\varrho\rangle\xrightarrow{\gamma(\nu,\mu)}\langle x'\parallel y', \varrho\rangle}$$

It must be pointed out that the communication function $\gamma(\nu,\mu)$ of two communicating actions $\nu$ and $\mu$ is used to exchange data between two interleaving quantum processes. Due to the quantum no-cloning theorem, the data must be exchanged by references (the names of the quantum variables), but not by values.

\textbf{Theorem \ref{AQCP}.1}. QPAP is a conservative extension of BQPA.

\begin{proof}
Since the corresponding TSS of BQPA is source-dependent, and the transition rules for merge operator $\parallel$, left merge operator $\leftmerge$ and communication merge $\mid$ contain only a fresh operator in their source, so the corresponding TSS of QPAP is a conservative extension of that of BQPA. That means that QPAP is a conservative extension of BQPA.
\end{proof}

\textbf{Theorem \ref{AQCP}.2}. Quantum bisimulation equivalence is a congruence with respenct to QPAP.

\begin{proof}
The structural part of QTSSs for QPAP and BQPA are all in panth format, so bisimulation equivalence that they induce is a congruence. According to the definition of quantum bisimulation, quantum bisimulation equivalence that QTSSs for QPAP induce is also a congruence.
\end{proof}

\subsubsection{Axiomatization for QPAP}

We design an axiomatization for QPAP illustrated in Table \ref{AxiomForQPAP}.

\begin{center}
\begin{table}
  \begin{tabular}{@{}ll@{}}
\hline No. &Axiom\\
  QM1 & $x\parallel y = (x\leftmerge y + y\leftmerge x) + x\mid y$ \\
  \\
  QLM2 & $\upsilon\leftmerge y = \upsilon\cdot y$ \\
  QLM3 & $(\upsilon\cdot x)\leftmerge y = \upsilon\cdot(x\parallel y)$ \\
  QLM4 & $(x + y)\leftmerge z = x\leftmerge z + y\leftmerge z$ \\
  \\
  QCM5 & $\nu\mid\mu=\gamma(\nu,\mu)$\\
  QCM6 & $\nu\mid(\mu\cdot y) = \gamma(\nu,\mu)\cdot y$\\
  QCM7 & $(\nu\cdot x)\mid\mu = \gamma(\nu,\mu)\cdot x$\\
  QCM8 & $(\nu\cdot x)\mid(\mu\cdot y) = \gamma(\nu,\mu)\cdot (x\parallel y)$\\
  QCM9 & $(x + y)\mid z = x\mid z + y\mid z$\\
  QCM10 & $x\mid (y+z) = x\mid y + x\mid z$\\
\end{tabular}
\caption{Axioms for QPAP}
\label{AxiomForQPAP}
\end{table}
\end{center}

Then, we can get the soundness and completeness theorems as follows.

\textbf{Theorem \ref{AQCP}.3}. $\mathcal{E}_{\textrm{QPAP}}$ is sound for QPAP modulo quantum bisimulation equivalence.

\begin{proof}
Since quantum bisimulation is both an equivalence and a congruence for QPAP, only the soundness of the first clause in the definition of the relation $=$ is needed to be checked. That is, if $s=t$ is an axiom in $\mathcal{E}_{\textrm{QPAP}}$ and $\sigma$ a closed substitution that maps the variable in $s$ and $t$ to basic quantum process terms, then we need to check that $\langle\sigma(s),\varrho\rangle\underline{\leftrightarrow}\langle\sigma(t),\varsigma\rangle$.

Since axioms in $\mathcal{E}_{\textrm{QPAP}}$ (same as $\mathcal{E}_{\textrm{PAP}}$) are sound for PAP modulo bisimulation equivalence, according to the definition of quantum bisimulation, we only need to check if $\varrho'=\varsigma'$ when $\varrho=\varsigma$, where $\varrho$ evolves into $\varrho'$ after execution of $\sigma(s)$ and $\varsigma$ evolves into $\varsigma'$ after execution of $\sigma(t)$. We can find that every axiom in Table \ref{AxiomForQPAP} meets the above condition.
\end{proof}

\textbf{Theorem \ref{AQCP}.4}. $\mathcal{E}_{\textrm{QPAP}}$ is complete for QPAP modulo quantum bisimulation equivalence.

\begin{proof}
To prove that $\mathcal{E}_{\textrm{QPAP}}$ is complete for QPAP modulo quantum bisilumation equivalence, it means that $\langle s, \varrho\rangle\underline{\leftrightarrow} \langle t,\varsigma\rangle$ implies $s=t$.

It was already proved that $\mathcal{E}_{\textrm{QPAP}}$ (same as $\mathcal{E}_{\textrm{PAP}}$) is complete for PAP modulo bisimulation equivalence, that is, $s\underline{\leftrightarrow} t$ implies $s=t$. $\langle s, \varrho\rangle\underline{\leftrightarrow} \langle t,\varsigma\rangle$ with $\varrho=\varsigma$ means that $s\underline{\leftrightarrow} t$ with $\varrho=\varsigma$ and $\varrho'=\varsigma'$, where $\varrho$ evolves into $\varrho'$ after execution of $s$ and $\varsigma$ evolves into $\varsigma'$ after execution of $t$, according to the definition of quantum bisimulation equivalence. The completeness of $\mathcal{E}_{\textrm{QPAP}}$ for PAP modulo bisimulation equivalence determines that $\mathcal{E}_{\textrm{QPAP}}$ is complete for QPAP modulo quantum bisimulation equivalence.
\end{proof}

\subsection{Deadlock and Encapsulation}

The mismatch of two communicating action pair $\nu$ and $\mu$ can cause a deadlock (nothing to do), we introduce the deadlock constant $\delta$ and extend the communication function $\gamma$ to $\gamma:C\times C\rightarrow C\cup\{\delta\}$. So, the introduction about communication merge $\mid$ in the above section should be with $\gamma(\nu,\mu)\neq \delta$. We also introduce a unary encapsulation operator $\partial_H$ for sets $H$ of atomic communicating actions, which rename all actions in $H$ into $\delta$. QPAP extended with deadlock constant $\delta$ and encapsulation operator $\partial_H$ is called the Algebra of Quantum Communicating Processes, which is abbreviated to AQCP.

\subsubsection{Transition Rules of AQCP}

The encapsulation operator $\partial_H(t)$ can execute all transitions of process term $t$ of which the labels are not in $H$, and does not change the quantum state, which is expressed by the following two transition rules.

$$\frac{\langle x, \varrho\rangle\xrightarrow{\nu}\langle\surd,\varrho\rangle}{\langle \partial_H(x),\varrho\rangle\xrightarrow{\nu}\langle \surd, \varrho\rangle}\quad\quad \nu\notin H$$

$$\frac{\langle x, \varrho\rangle\xrightarrow{\nu}\langle x',\varrho\rangle}{\langle \partial_H(x),\varrho\rangle\xrightarrow{\nu}\langle \partial_H(x'), \varrho\rangle}\quad\quad \nu\notin H$$

\textbf{Theorem \ref{AQCP}.5}. AQCP is a conservative extension of QPAP.

\begin{proof}
Since the corresponding TSS of QPAP is source-dependent, and the transition rules for encapsulation operator $\partial_H$ contain only a fresh operator in their source, so the corresponding TSS of AQCP is a conservative extension of that of QPAP. That means that AQCP is a conservative extension of QPAP.
\end{proof}

\textbf{Theorem \ref{AQCP}.6}. Quantum bisimulation equivalence is a congruence with respect to AQCP.

\begin{proof}
The structural part of QTSSs for AQCP and QPAP are all in panth format, so bisimulation equivalence that they induce is a congruence. According to the definition of quantum bisimulation, quantum bisimulation equivalence that QTSSs for AQCP induce is also a congruence.
\end{proof}

\subsubsection{Axiomatization for AQCP}

The axioms for AQCP are shown in Table \ref{AxiomForAQCP}.

\begin{center}
\begin{table}
  \begin{tabular}{@{}ll@{}}
\hline No. &Axiom\\
  QA6 & $x+\delta = x$ \\
  QA7 & $\delta\cdot x = \delta$ \\
  \\
  QD1 & $\nu\notin H\quad \partial_H(\nu) = \nu$ \\
  QD2 & $\nu\in H\quad \partial_H(\nu) = \delta$ \\
  QD3 & $\partial_H(\delta) = \delta$\\
  QD4 & $\partial_H(x+y)=\partial_H(x)+\partial_H(y)$\\
  QD5 & $\partial_H(x\cdot y)=\partial_H(x)\cdot\partial_H(y)$\\
  \\
  QLM11 & $\delta\leftmerge x=\delta$\\
  QCM12 & $\delta\mid x = \delta$\\
  QCM13 & $x\mid\delta = \delta$\\
\end{tabular}
\caption{Axioms for AQCP}
\label{AxiomForAQCP}
\end{table}
\end{center}

The soundness and completeness theorems are following.

\textbf{Theorem \ref{AQCP}.7}. $\mathcal{E}_{\textrm{AQCP}}$ is sound for AQCP modulo quantum bisimulation equivalence.

\begin{proof}
Since quantum bisimulation is both an equivalence and a congruence for AQCP, only the soundness of the first clause in the definition of the relation $=$ is needed to be checked. That is, if $s=t$ is an axiom in $\mathcal{E}_{\textrm{AQCP}}$ and $\sigma$ a closed substitution that maps the variable in $s$ and $t$ to basic quantum process terms, then we need to check that $\langle\sigma(s),\varrho\rangle\underline{\leftrightarrow}\langle\sigma(t),\varsigma\rangle$.

Since axioms in $\mathcal{E}_{\textrm{AQCP}}$ (same as $\mathcal{E}_{\textrm{ACP}}$) are sound for ACP modulo bisimulation equivalence, according to the definition of quantum bisimulation, we only need to check if $\varrho'=\varsigma'$ when $\varrho=\varsigma$, where $\varrho$ evolves into $\varrho'$ after execution of $\sigma(s)$ and $\varsigma$ evolves into $\varsigma'$ after execution of $\sigma(t)$. We can find that every axiom in Table \ref{AxiomForAQCP} meets the above condition.
\end{proof}

\textbf{Theorem \ref{AQCP}.8}. $\mathcal{E}_{\textrm{AQCP}}$ is complete for AQCP modulo quantum bisimulation equivalence.

\begin{proof}
To prove that $\mathcal{E}_{\textrm{AQCP}}$ is complete for AQCP modulo quantum bisilumation equivalence, it means that $\langle s, \varrho\rangle\underline{\leftrightarrow} \langle t,\varsigma\rangle$ implies $s=t$.

It was already proved that $\mathcal{E}_{\textrm{AQCP}}$ (same as $\mathcal{E}_{\textrm{ACP}}$) is complete for ACP modulo bisimulation equivalence, that is, $s\underline{\leftrightarrow} t$ implies $s=t$. $\langle s, \varrho\rangle\underline{\leftrightarrow} \langle t,\varsigma\rangle$ with $\varrho=\varsigma$ means that $s\underline{\leftrightarrow} t$ with $\varrho=\varsigma$ and $\varrho'=\varsigma'$, where $\varrho$ evolves into $\varrho'$ after execution of $s$ and $\varsigma$ evolves into $\varsigma'$ after execution of $t$, according to the definition of quantum bisimulation equivalence. The completeness of $\mathcal{E}_{\textrm{AQCP}}$ for ACP modulo bisimulation equivalence determines that $\mathcal{E}_{\textrm{AQCP}}$ is complete for AQCP modulo quantum bisimulation equivalence.
\end{proof}

\section{Recursion}\label{Recursion}

To capture infinite computing, recursion is introduced in this section. In the following, $E,F,G$ are guarded linear recursion specifications, $X,Y,Z$ are recursive variables. We first introduce several important concepts, which come from \cite{ACP}.

\textbf{Definition \ref{Recursion}.1 (Recursive specification)}. A recursive specification is a finite set of recursive equations

$$X_1=t_1(X_1,\cdots,X_n)$$
$$...$$
$$X_n=t_n(X_1,\cdots,X_n)$$

where the left-hand sides of $X_i$ are called recursion variables, and the right-hand sides $t_i(X_1,\cdots,X_n)$ are process terms in AQCP with possible occurrences of the recursion variables $X_1,\cdots,X_n$.

\textbf{Definition \ref{Recursion}.2 (Solution)}. Processes $p_1,\cdots,p_n$ are a solution for a recursive specification $\{X_i=t_i(X_1,\cdots,X_n)|i\in\{1,\cdots,n\}\}$ (with respect to bisimulation equivalence) if $p_i\underline{\leftrightarrow}t_i(p_1,\cdots,p_n)$ for $i\in\{1,\cdots,n\}$.

\textbf{Definition \ref{Recursion}.3 (Guarded recursive specification)}. A recursive specification

$$X_1=t_1(X_1,\cdots,X_n)$$
$$...$$
$$X_n=t_n(X_1,\cdots,X_n)$$

is guarded if the right-hand sides of its recursive equations can be adapted to the form by applications of the axioms in $\mathcal{E}_{\textrm{AQCP}}$ and replacing recursion variables by the right-hand sides of their recursive equations,

$$\alpha_1\cdot s_1(X_1,\cdots,X_n)+\cdots+\alpha_k\cdot s_k(X_1,\cdots,X_n)+\beta_1+\cdots+\beta_l$$

where $\alpha_1,\cdots,\alpha_k,\beta_1,\cdots,\beta_l\in A\cup C$, and the sum above is allowed to be empty, in which case it represents the deadlock $\delta$.

\textbf{Definition \ref{Recursion}.4 (Linear recursive specification)}. A recursive specification is linear if its recursive equations are of the form

$$\alpha_1X_1+\cdots+\alpha_kX_k+\beta_1+\cdots+\beta_l$$

where $\alpha_1,\cdots,\alpha_k,\beta_1,\cdots,\beta_l\in A\cup C$, and the sum above is allowed to be empty, in which case it represents the deadlock $\delta$.

\subsection{Transition Rules of Guarded Recursion}

For a guarded recursive specifications $E$ with the form

$$X_1=t_1(X_1,\cdots,X_n)$$
$$\cdots$$
$$X_n=t_n(X_1,\cdots,X_n)$$

the behavior of the solution $\langle X_i|E$ for the recursion variable $X_i$ in $E$, where $i\in\{1,\cdots,n\}$, is exactly the behavior of their right-hand sides $t_i(X_1,\cdots,X_n)$, which is captured by the following two transition rules.

$$\frac{\langle t_i(\langle X_1|E\rangle,\cdots,\langle X_n|E\rangle),\varrho\rangle\xrightarrow{\upsilon}\langle\surd,\varrho'\rangle}{\langle\langle X_i|E\rangle,\varrho\rangle\xrightarrow{\upsilon}\langle\surd,\varrho'\rangle}$$

$$\frac{\langle t_i(\langle X_1|E\rangle,\cdots,\langle X_n|E\rangle),\varrho\rangle\xrightarrow{\upsilon}\langle y,\varrho'\rangle}{\langle\langle X_i|E\rangle,\varrho\rangle\xrightarrow{\upsilon}\langle y,\varrho'\rangle}$$

\textbf{Theorem \ref{Recursion}.1}. AQCP with guarded recursion is a conservative extension of AQCP.

\begin{proof}
Since the corresponding TSS of AQCP is source-dependent, and the transition rules for guarded recursion contain only a fresh constant in their source, so the corresponding TSS of AQCP with guarded recursion is a conservative extension of that of AQCP. That means that AQCP with guarded recursion is a conservative extension of AQCP.
\end{proof}

\textbf{Theorem \ref{Recursion}.2}. Quantum bisimulation equivalence is a congruence with respect to AQCP with guarded recursion.

\begin{proof}
The structural part of QTSSs for guarded recursion and AQCP are all in panth format, so bisimulation equivalence that they induce is a congruence. According to the definition of quantum bisimulation, quantum bisimulation equivalence that QTSSs for AQCP with guarded recursion induce is also a congruence.
\end{proof}

\subsection{Axiomatization for Guarded Recursion}

The RDP (Recursive Definition Principle) and the RSP (Recursive Specification Principle) are shown in Table \ref{RDPAndRSP}.

\begin{center}
\begin{table}
  \begin{tabular}{@{}ll@{}}
\hline No. &Axiom\\
  RDP & $\langle X_i|E\rangle = t_i(\langle X_1|E,\cdots,X_n|E\rangle)\quad\quad (i\in\{1,\cdots,n\})$\\
  RSP & if $y_i=t_i(y_1,\cdots,y_n)$ for $i\in\{1,\cdots,n\}$, then $y_i=\langle X_i|E\rangle \quad\quad (i\in\{1,\cdots,n\})$\\
\end{tabular}
\caption{Recursive definition principle and recursive specification principle}
\label{RDPAndRSP}
\end{table}
\end{center}

\textbf{Theorem \ref{Recursion}.3}. $\mathcal{E}_{\textrm{AQCP}}$ + RDP + RSP is sound for AQCP with guarded recursion modulo quantum bisimulation equivalence.

\begin{proof}
Since quantum bisimulation is both an equivalence and a congruence for AQCP with guarded recursion, only the soundness of the first clause in the definition of the relation $=$ is needed to be checked. That is, if $s=t$ is an axiom in $\mathcal{E}_{\textrm{AQCP}}$ + RDP + RSP and $\sigma$ a closed substitution that maps the variable in $s$ and $t$ to basic quantum process terms, then we need to check that $\langle\sigma(s),\varrho\rangle\underline{\leftrightarrow}\langle\sigma(t),\varsigma\rangle$.

Since axioms in $\mathcal{E}_{\textrm{AQCP}}$ + RDP + RSP (same as $\mathcal{E}_{\textrm{ACP}}$ + RDP + RSP) are sound for ACP with guarded recursion modulo bisimulation equivalence, according to the definition of quantum bisimulation, we only need to check if $\varrho'=\varsigma'$ when $\varrho=\varsigma$, where $\varrho$ evolves into $\varrho'$ after execution of $\sigma(s)$ and $\varsigma$ evolves into $\varsigma'$ after execution of $\sigma(t)$. We can find that every axiom in Table \ref{RDPAndRSP} meets the above condition.
\end{proof}

\textbf{Theorem \ref{Recursion}.4}. $\mathcal{E}_{\textrm{AQCP}}$ + RDP + RSP is complete for AQCP with linear recursion modulo quantum bisimulation equivalence.

\begin{proof}
To prove that $\mathcal{E}_{\textrm{AQCP}}$ + RDP + RSP is complete for AQCP with linear recursion modulo quantum bisilumation equivalence, it means that $\langle s, \varrho\rangle\underline{\leftrightarrow} \langle t,\varsigma\rangle$ implies $s=t$.

It was already proved that $\mathcal{E}_{\textrm{AQCP}}$ + RDP + RSP (same as $\mathcal{E}_{\textrm{ACP}}$ + RDP +RSP) is complete for ACP with linear recursion modulo bisimulation equivalence, that is, $s\underline{\leftrightarrow} t$ implies $s=t$. $\langle s, \varrho\rangle\underline{\leftrightarrow} \langle t,\varsigma\rangle$ with $\varrho=\varsigma$ means that $s\underline{\leftrightarrow} t$ with $\varrho=\varsigma$ and $\varrho'=\varsigma'$, where $\varrho$ evolves into $\varrho'$ after execution of $s$ and $\varsigma$ evolves into $\varsigma'$ after execution of $t$, according to the definition of quantum bisimulation equivalence. The completeness of $\mathcal{E}_{\textrm{AQCP}}$ + RDP + RSP for ACP with linear recursion modulo bisimulation equivalence determines that $\mathcal{E}_{\textrm{AQCP}}$ + RDP + RSP is complete for AQCP with linear recursion modulo quantum bisimulation equivalence.
\end{proof}

\section{Abstraction}\label{Abstraction}

A quantum program has internal implementations and external behaviors. Abstraction technology abstracts away from the internal steps to check if the internal implementations really display the desired external behaviors. This makes the introduction of special silent step constant $\tau$ and the abstraction operator $\tau_I$.

Firstly, we introduce the concept of guarded linear recursive specification, which comes from \cite{ACP}.

\textbf{Definition \ref{Abstraction}.1 (Guarded linear recursive specification)}. A recursive specification is linear if its recursive equations are of the form

$$\alpha_1X_1+\cdots+\alpha_kX_k+\beta_1+\cdots+\beta_l$$

where $\alpha_1,\cdots,\alpha_k,\beta_1,\cdots,\beta_l\in A\cup C\cup\{\tau\}$.

A linear recursive specification $E$ is guarded if there does not exist an infinite sequence of $\tau$-transitions $\langle X|E\rangle\xrightarrow{\tau}\langle X'|E\rangle\xrightarrow{\tau}\langle X''|E\rangle\xrightarrow{\tau}\cdots$.

\subsection{Silent Step}

A $\tau$-transition is silent, which is means that it can be eliminated from a quantum process graph. $\tau$ is an internal step and keep silent from an external observer, but please remember, $\tau$ is a quantum operation in nature. This fact makes that $\tau$ must influence the state of all quantum variables $\varrho$, that is, $\tau$ is not really silent for a quantum process configuration $\langle p,\varrho\rangle$. To make $\tau$ keep silent, the definition of $\varrho$ must be changed, that is, $\varrho$ does not record the state of all quantum variables, some variables must be moved away. But, what variables should be moved away? The quantum variables that $\tau$ may influence are called private variables. These private variables include not only the variables $\tau$ directly manipulates, but also those variables which are entangled with the variables that $\tau$ directly manipulates. The quantum variables that $\tau$ can not influence are called public variables. In the following, $\varrho$ records the state of all public variables. We use the symbol $\tau(\varrho)$ to denote the state of all public quantum variables after execution of $\tau$. From an external view, we can see that $\varrho = \tau(\varrho)$.

The processing of $\tau$ in quantum processes is some what farfetched. But, it is the only choice under the framework of quantum process configuration $\langle p,\varrho\rangle$. Otherwise, the concept of branching bisimulation (weak bisimularity) and the theory of abstraction can not be established.

Now, the set $A$ of all quantum operations is extended to $A\cup\{\tau\}$, $C$ to $C\cup\{\tau\}$, and $\gamma$ to $\gamma:C\cup\{\tau\}\times C\cup\{\tau\}\rightarrow C\cup\{\delta\}$.

\subsubsection{Transition Rules of Silent Step}

$\tau$ keeps silent from an external observer, which is expressed by the following transition rules.

$$\frac{}{\langle \tau, \varrho\rangle\xrightarrow{\tau}\langle\surd,\tau(\varrho)\rangle}$$

Transition rules for alternative composition, sequential composition and guarded linear recursion that involves $\tau$-transitions are omitted.

\textbf{Theorem \ref{Abstraction}.1}. AQCP with silent step and guarded linear recursion is a conservative extension of AQCP with guarded linear recursion.

\begin{proof}
The corresponding TSS of AQCP with silent step and guarded linear recursion is a conservative extension of that of AQCP with guarded linear recursion. That means that AQCP with silent step and guarded linear recursion is a conservative extension of AQCP with guarded linear recursion.
\end{proof}

\textbf{Theorem \ref{Abstraction}.2}. Quantum rooted branching bisimulation equivalence is a congruence with respect to AQCP with silent step and guarded linear recursion.

\begin{proof}
The structural part of QTSSs for AQCP with silent step and guarded linear recursion are all in RBB cool format by incorporating the successful termination predicate $\downarrow$ in the transition rules, so rooted branching bisimulation equivalence that they induce is a congruence. According to the definition of quantum rooted branching bisimulation, quantum rooted branching bisimulation equivalence that QTSSs for AQCP with silent step and guarded linear recursion induce is also a congruence.
\end{proof}

\subsubsection{Axioms for Silent Step}

The axioms for silent step is shown in Table \ref{AxiomForSS}.

\begin{center}
\begin{table}
  \begin{tabular}{@{}ll@{}}
\hline No. &Axiom\\
  QB1 & $\upsilon\cdot\tau= \upsilon$ \\
  QB2 & $\upsilon\cdot(\tau\cdot(x+y)+x)=\upsilon\cdot(x+y)$ \\
\end{tabular}
\caption{Axioms for silent step}
\label{AxiomForSS}
\end{table}
\end{center}

\textbf{Theorem \ref{Abstraction}.3}. $\mathcal{E}_{\textrm{AQCP}}$ + QB1,QB2 + RDP + RSP is sound for AQCP with silent step and guarded linear recursion, modulo quantum rooted branching bisimulation equivalence.

\begin{proof}
Since quantum rooted branching bisimulation is both an equivalence and a congruence for ACP with silent step and guarded linear recursion, only the soundness of the first clause in the definition of the relation $=$ is needed to be checked. That is, if $s=t$ is an axiom in $\mathcal{E}_{\textrm{AQCP}}$ + QB1,QB2 + RDP + RSP and $\sigma$ a closed substitution that maps the variable in $s$ and $t$ to basic quantum process terms, then we need to check that $\langle\sigma(s),\varrho\rangle\underline{\leftrightarrow}_{rb}\langle\sigma(t),\varsigma\rangle$.

Since axioms in $\mathcal{E}_{\textrm{AQCP}}$ + QB1,QB2 + RDP + RSP (same as $\mathcal{E}_{\textrm{ACP}}$ + QB1,QB2 + RDP + RSP) are sound for ACP with silent step and guarded linear recursion modulo rooted branching bisimulation equivalence, according to the definition of quantum rooted branching bisimulation, we only need to check if $\varrho'=\varsigma'$ when $\varrho=\varsigma$, where $\varrho$ evolves into $\varrho'$ after execution of $\sigma(s)$ and $\varsigma$ evolves into $\varsigma'$ after execution of $\sigma(t)$. We can find that every axiom in Table \ref{AxiomForSS} meets the above condition.
\end{proof}

\textbf{Theorem \ref{Abstraction}.4}. $\mathcal{E}_{\textrm{AQCP}}$ + QB1,QB2 + RDP + RSP is complete for AQCP with silent step and guarded linear recursion, modulo quantum rooted branching bisimulation equivalence.

\begin{proof}
To prove that $\mathcal{E}_{\textrm{AQCP}}$ + QB1,QB2 + RDP + RSP is complete for AQCP with silent step and guarded linear recursion modulo quantum rooted branching bisilumation equivalence, it means that $\langle s, \varrho\rangle\underline{\leftrightarrow}_{rb} \langle t,\varsigma\rangle$ implies $s=t$.

It was already proved that $\mathcal{E}_{\textrm{AQCP}}$ + QB1,QB2 + RDP + RSP (same as $\mathcal{E}_{\textrm{ACP}}$ + QB1,QB2 + RDP + RSP) is complete for ACP with silent step and guarded linear recursion modulo rooted branching bisimulation equivalence, that is, $s\underline{\leftrightarrow}_{rb} t$ implies $s=t$. $\langle s, \varrho\rangle\underline{\leftrightarrow}_{rb} \langle t,\varsigma\rangle$ with $\varrho=\varsigma$ means that $s\underline{\leftrightarrow}_{rb} t$ with $\varrho=\varsigma$ and $\varrho'=\varsigma'$, where $\varrho$ evolves into $\varrho'$ after execution of $s$ and $\varsigma$ evolves into $\varsigma'$ after execution of $t$, according to the definition of quantum rooted branching bisimulation equivalence. The completeness of $\mathcal{E}_{\textrm{AQCP}}$ + QB1,QB2 + RDP + RSP for ACP with silent step and guarded linear recursion modulo rooted branching bisimulation equivalence determines that $\mathcal{E}_{\textrm{AQCP}}$ + QB1,QB2 + RDP + RSP is complete for AQCP with silent step and guarded linear recursion modulo quantum rooted branching bisimulation equivalence.
\end{proof}

\subsection{Abstraction}

Abstraction operator $\tau_I$ is used to abstract away the internal implementations. AQCP extended with silent step $\tau$ and abstraction operator $\tau_I$ is denoted by $\textrm{AQCP}_{\tau}$.

\subsubsection{Transition Rules of Abstraction Operator}

Abstraction operator $\tau_I(t)$ renames all labels of transitions of $t$ that are in the set $I$ into $\tau$, and does not change the state of all public quantum variables, which is captured by the following four transition rules.

$$\frac{\langle x, \varrho\rangle\xrightarrow{\upsilon}\langle\surd, \varrho'\rangle}{\langle\tau_I(x),\varrho\rangle\xrightarrow{\upsilon}\langle\surd,\varrho'\rangle}\quad \upsilon\notin I$$

$$\frac{\langle x, \varrho\rangle\xrightarrow{\upsilon}\langle x', \varrho'\rangle}{\langle\tau_I(x),\varrho\rangle\xrightarrow{\upsilon}\langle\tau_I(x'),\varrho'\rangle}\quad \upsilon\notin I$$

$$\frac{\langle x, \varrho\rangle\xrightarrow{\upsilon}\langle\surd, \varrho'\rangle}{\langle\tau_I(x),\varrho\rangle\xrightarrow{\tau}\langle\surd,\tau(\varrho)\rangle}\quad \upsilon\in I$$

$$\frac{\langle x, \varrho\rangle\xrightarrow{\upsilon}\langle x', \varrho'\rangle}{\langle\tau_I(x),\varrho\rangle\xrightarrow{\tau}\langle\tau_I(x'),\tau(\varrho)\rangle}\quad \upsilon\in I$$

Note that $\varrho=\tau(\varrho)=\tau(\varrho')$ in the sense of public variables.

\textbf{Theorem \ref{Abstraction}.5}. $\textrm{AQCP}_{\tau}$ with guarded linear recursion is a conservative extension of AQCP with silent step and guarded linear recursion.

\begin{proof}
The corresponding TSS of $\textrm{AQCP}_{\tau}$ guarded linear recursion is a conservative extension of that of AQCP with silent step and guarded linear recursion. That means that $\textrm{AQCP}_{\tau}$ guarded linear recursion is a conservative extension of AQCP with silent step and guarded linear recursion.
\end{proof}

\textbf{Theorem \ref{Abstraction}.6}. Quantum rooted branching bisimulation equivalence is a congruence with respect to $\textrm{AQCP}_{\tau}$ with guarded linear recursion.

\begin{proof}
The structural part of QTSSs for $\textrm{AQCP}_{\tau}$ guarded linear recursion are all in RBB cool format by incorporating the successful termination predicate $\downarrow$ in the transition rules, so rooted branching bisimulation equivalence that they induce is a congruence. According to the definition of quantum rooted branching bisimulation, quantum rooted branching bisimulation equivalence that QTSSs for $\textrm{AQCP}_{\tau}$ guarded linear recursion induce is also a congruence.
\end{proof}

\subsubsection{Axiomatization for Abstraction Operator}

The axioms for abstraction operator are shown in Table \ref{AxiomForAO}.

\begin{center}
\begin{table}
  \begin{tabular}{@{}ll@{}}
\hline No. &Axiom\\
  QTI1 & $\upsilon\notin I \quad \tau_I(\upsilon)=\upsilon$\\
  QTI2 & $\upsilon\in I \quad \tau_I(\upsilon)=\tau$\\
  QTI3 & $\tau_I(\delta)=\delta$\\
  QTI4 & $\tau_I(x+y)=\tau_I(x)+\tau_I(y)$\\
  QTI5 & $\tau_I(x\cdot y)=\tau_I(x)\cdot\tau_I(y)$\\
\end{tabular}
\caption{Axioms for abstraction operator}
\label{AxiomForAO}
\end{table}
\end{center}

Before we introduce the cluster fair abstraction rule, the concept of cluster is given firstly, which comes from \cite{ACP}.

\textbf{Definition \ref{Abstraction}.2 (Cluster)}. Let $E$ be a guarded linear recursive specification, and $I\subseteq A$. Two recursion variable $X$ and $Y$ in $E$ are in the same cluster for $I$ if and only if there exist sequences of transitions $\langle X|E\rangle\xrightarrow{\beta_1}\cdots\xrightarrow{\beta_m}\langle Y|E\rangle$ and $\langle Y|E\rangle\xrightarrow{\eta_1}\cdots\xrightarrow{\eta_n}\langle X|E\rangle$, where $\beta_1,\cdots,\beta_m,\eta_1,\cdots,\eta_n\in I\cup\{\tau\}$.

$\alpha$ or $\alpha X$ is an exit for the cluster $C$ if and only if: (1) $\alpha$ or $\alpha X$ is a summand at the right-hand side of the recursive equation for a recursion variable in $C$, and (2) in the case of $\alpha X$, either $\alpha\notin I\cup\{\tau\}$ or $X\notin C$.

\begin{center}
\begin{table}
  \begin{tabular}{@{}ll@{}}
\hline No. &Axiom\\
  CFAR & If $X$ is in a cluster for $I$ with exits $\{\upsilon_1Y_1,\cdots,\upsilon_mY_m,\omega_1,\cdots,\omega_n\}$, \\ & then $\tau\cdot\tau_I(\langle X|E\rangle)=\tau\cdot\tau_I(\upsilon_1\langle Y_1|E\rangle,\cdots,\upsilon_m\langle Y_m|E\rangle,\omega_1,\cdots,\omega_n)$\\
\end{tabular}
\caption{Cluster fair abstraction rule}
\label{CFAR}
\end{table}
\end{center}

\textbf{Theorem \ref{Abstraction}.7}. $\mathcal{E}_{\textrm{AQCP}_{\tau}}$ + RSP + RDP + CFAR is sound for $\textrm{AQCP}_{\tau}$ with guarded linear recursion, modulo quantum rooted branching bisimulation equivalence.

\begin{proof}
Since quantum rooted branching bisimulation is both an equivalence and a congruence for $\textrm{AQCP}_{\tau}$ guarded linear recursion, only the soundness of the first clause in the definition of the relation $=$ is needed to be checked. That is, if $s=t$ is an axiom in $\mathcal{E}_{\textrm{AQCP}_{\tau}}$ + RSP + RDP + CFAR and $\sigma$ a closed substitution that maps the variable in $s$ and $t$ to basic quantum process terms, then we need to check that $\langle\sigma(s),\varrho\rangle\underline{\leftrightarrow}_{rb}\langle\sigma(t),\varsigma\rangle$.

Since axioms in $\mathcal{E}_{\textrm{AQCP}_{\tau}}$ + RSP + RDP + CFAR (same as $\mathcal{E}_{\textrm{ACP}_{\tau}}$ + RSP + RDP + CFAR) are sound for $\textrm{ACP}_{\tau}$ with guarded linear recursion modulo rooted branching bisimulation equivalence, according to the definition of quantum rooted branching bisimulation, we only need to check if $\varrho'=\varsigma'$ when $\varrho=\varsigma$, where $\varrho$ evolves into $\varrho'$ after execution of $\sigma(s)$ and $\varsigma$ evolves into $\varsigma'$ after execution of $\sigma(t)$. We can find that every axiom in Table \ref{AxiomForAO} and Table \ref{CFAR} meets the above condition.
\end{proof}

\textbf{Theorem \ref{Abstraction}.8}. $\mathcal{E}_{\textrm{AQCP}_{\tau}}$ + RSP + RDP + CFAR is complete for $\textrm{AQCP}_{\tau}$ with guarded linear recursion, modulo quantum rooted branching bisimulation equivalence.

\begin{proof}
To prove that $\mathcal{E}_{\textrm{AQCP}_{\tau}}$ + RSP + RDP + CFAR is complete for $\textrm{AQCP}_{\tau}$ with guarded linear recursion modulo quantum rooted branching bisilumation equivalence, it means that $\langle s, \varrho\rangle\underline{\leftrightarrow}_{rb} \langle t,\varsigma\rangle$ implies $s=t$.

It was already proved that $\mathcal{E}_{\textrm{AQCP}_{\tau}}$ + RSP + RDP + CFAR (same as $\mathcal{E}_{\textrm{ACP}_{\tau}}$ + RSP + RDP + CFAR) is complete for $\textrm{ACP}_{\tau}$ with guarded linear recursion modulo rooted branching bisimulation equivalence, that is, $s\underline{\leftrightarrow}_{rb} t$ implies $s=t$. $\langle s, \varrho\rangle\underline{\leftrightarrow}_{rb} \langle t,\varsigma\rangle$ with $\varrho=\varsigma$ means that $s\underline{\leftrightarrow}_{rb} t$ with $\varrho=\varsigma$ and $\varrho'=\varsigma'$, where $\varrho$ evolves into $\varrho'$ after execution of $s$ and $\varsigma$ evolves into $\varsigma'$ after execution of $t$, according to the definition of quantum rooted branching bisimulation equivalence. The completeness of $\mathcal{E}_{\textrm{AQCP}_{\tau}}$ + RSP + RDP + CFAR for $\textrm{ACP}_{\tau}$ with guarded linear recursion modulo rooted branching bisimulation equivalence determines that $\mathcal{E}_{\textrm{AQCP}_{\tau}}$ + RSP + RDP + CFAR is complete for $\textrm{AQCP}_{\tau}$ with guarded linear recursion modulo quantum rooted branching bisimulation equivalence.
\end{proof}

\section{Unifying Quantum and Classical Computing}\label{Unifying}

We use a quantum process configurations $\langle p,\varrho\rangle$ to represent information related to the execution of a quantum process, in which $p$ represents the structural properties of a quantum process and $\varrho$ expresses the quantum properties of a quantum process. We have established a whole theory about quantum processes based on ACP, which is called qACP.

In qACP, the set $A$ of actions is consisted of atomic quantum operations, and also the deadlock $\delta$ and the silent step $\tau$. The execution of an atomic quantum operation $\alpha$ not only influences of the structural part $p$, but also changes the state of quantum variables $\varrho$. We still use the framework of a quantum process configuration $p,\varrho$ under the situation of classical computing. In classical computing, the execution of a (classical) atomic action $a$ only influence the structural part $p$, and maintain the quantum state $\varrho$ unchanged. Note that, this kind of actions are already introduced in AQCP in section \ref{AQCP}, which are called quantum communicating actions, and range over the set $C$ of quantum communicating actions. In nature, quantum communicating actions are some kind of classical actions in contrast to quantum operations, because they are unrelated to the quantum state $\varrho$. The difference of a quantum communicating action and a classical communicating action is that they exchange different contents, a classical communicating action exchange the classical data by value or by reference, while a quantum communicating action exchange the quantum variables only by reference. We extend the set $C$ of quantum communicating actions to classical atomic actions (including classical communicating actions), and variables $\nu,\mu$ range over $C$, and $a,b\in C$.

Base on the fact that a classical action $a$ does not affect the quantum state $\varrho$, we can generalize classical ACP under the framework of quantum process configuration $\langle p, \varrho\rangle$. We only take an example of BPA, while PAP, ACP, ACP with guarded linear recursion, $\textrm{ACP}_{\tau}$ with guarded linear recursion are omitted.

We give the transition rules under quantum transition system specification (QTSS) for BPA as follows.

$$\frac{}{\langle\upsilon,\varrho\rangle\xrightarrow{\nu}\langle\surd,\varrho\rangle}$$

$$\frac{\langle x,\varrho\rangle\xrightarrow{\nu}\langle\surd,\varrho\rangle}{\langle x+y,\varrho\rangle\xrightarrow{\nu}\langle\surd,\varrho\rangle}$$

$$\frac{\langle x,\varrho\rangle\xrightarrow{\nu}\langle x',\varrho\rangle}{\langle x+y,\varrho\rangle\xrightarrow{\nu}\langle x',\varrho\rangle}$$

$$\frac{\langle y,\varrho\rangle\xrightarrow{\nu}\langle\surd,\varrho\rangle}{\langle x+y,\varrho\rangle\xrightarrow{\nu}\langle\surd,\varrho\rangle}$$

$$\frac{\langle y,\varrho\rangle\xrightarrow{\nu}\langle y',\varrho\rangle}{\langle x+y,\varrho\rangle\xrightarrow{\nu}\langle y',\varrho\rangle}$$

$$\frac{\langle x,\varrho\rangle\xrightarrow{\nu}\langle\surd,\varrho\rangle}{\langle x\cdot y,\varrho\rangle\xrightarrow{\nu}\langle y,\varrho\rangle}$$

$$\frac{\langle x,\varrho\rangle\xrightarrow{\nu}\langle x',\varrho\rangle}{\langle x\cdot y,\varrho\rangle\xrightarrow{\nu}\langle x'\cdot y,\varrho\rangle}$$

We design an axiomatization $\mathcal{E}_{\textrm{BPA}}$ for BPA modulo quantum bisimulation equivalence as Table \ref{AxiomForBPA} shows.

\begin{center}
\begin{table}
  \begin{tabular}{@{}ll@{}}
\hline No. &Axiom\\
  A1 & $x + y = y + x$ \\
  A2 & $(x + y) + z = x + (y + z)$ \\
  A3 & $x + x = x$ \\
  A4 & $(x + y)\cdot z = x\cdot z + y\cdot z$ \\
  A5 & $(x\cdot y)\cdot z = x\cdot (y\cdot z)$\\
\end{tabular}
\caption{Axioms for BPA}
\label{AxiomForBPA}
\end{table}
\end{center}

We can get the following conclusions naturally.

\textbf{Theorem \ref{Unifying}.1}. Quantum bisimulation equivalence is a congruence with respect to BPA.

\textbf{Theorem \ref{Unifying}.2}. $\mathcal{E}_{\textrm{BPA}}$ is sound for BPA modulo quantum bisimulation equivalence.

\textbf{Theorem \ref{Unifying}.3}. $\mathcal{E}_{\textrm{BPA}}$ is complete for BPA modulo quantum bisimulation equivalence.

Note that, the behavior of deadlock constant $\delta$ quantum computing is the same as that of classical computing. But, the behavior of silent step $\tau$ is different under the framework of quantum process configuration $\langle p, \varrho\rangle$ for quantum computing and classical computing, just because $\tau$ in quantum computing can affect the state of all quantum variables, while $\tau$ in classical computing really keeps silent.

Making classical ACP (including BPA, PAP, ACP, ACP with guarded linear recursion, and $\textrm{ACP}_{\tau}$ with guarded linear recursion) being under the framework of quantum process configuration $\langle p,\varrho\rangle$ for classical computing is trivial, because $\varrho$ is meaningless only for classical computing. But, in the view of unifying quantum computing and classical computing, this work would be very important. Fortunately, qACP and classical ACP are unified under the framework of quantum process configuration $\langle p, \varrho\rangle$, that is, qACP and classical ACP have the same equational logic (axiomatization $\mathcal{E}_{\textrm{qACP}}$ and $\mathcal{E}_{\textrm{ACP}}$) and the same semantic model (strong quantum bisimularity and weak quantum bisimularity).

The unifying of qACP and classical ACP has an important significance, because most quantum protocols, like the famous BB84 protocol\cite{BB84}, are mixtures of quantum information and classical information, and those of quantum computing and classical computing. This unifying can be used widely in verification for all quantum protocols.

\section{Verification for Quantum Protocols -- The BB84 Protocol}\label{Verification}

The unifying of qACP and classical ACP under the framework of quantum process configuration $\langle p,\varrho\rangle$ makes verification for quantum protocols possible, not only the pure quantum protocol, but also protocol that mixes quantum information and classical information.

The famous BB84 protocol\cite{BB84} is the first quantum key distribution protocol, in which quantum information and classical information are mixed. We take an example of the BB84 protocol to illustrate the usage of qACP in verification of quantum protocols.

\begin{figure}
  \centering
  \includegraphics{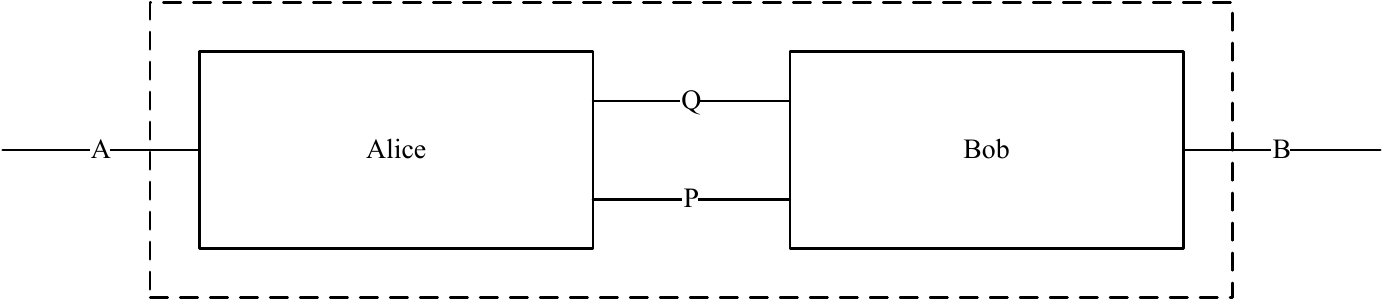}
  \caption{The BB84 protocol.}
  \label{BB84}
\end{figure}

The BB84 protocol is used to create a private key between two parities, Alice and Bob. Firstly, we introduce the basic BB84 protocol briefly, which is illustrated in Fig.\ref{BB84}.

\begin{enumerate}
  \item Alice create two string of bits with size $n$ randomly, denoted as $B_a$ and $K_a$.
  \item Alice generates a string of qubits $q$ with size $n$, and the $i$th qubit is $q$ is $|x_y\rangle$, where $x$ is the $i$th bit of $B_a$ and $y$ is the $i$th bit of $K_a$.
  \item Alice sends $q$ to Bob through a quantum channel $Q$ between Alice and Bob.
  \item Bob receives $q$ and randomly generates a string of bits $B_b$ with size $n$.
  \item Bob measures each qubit of $q$ according to a basis by bits of $B_b$. And the measurement results would be $K_b$, which is also with size $n$.
  \item Bob sends his measurement bases $B_b$ to Alice through a public channel $P$.
  \item Once receiving $B_b$, Alice sends her bases $B_a$ to Bob through channel $P$, and Bob receives $B_a$.
  \item Alice and Bob determine that at which position the bit strings $B_a$ and $B_b$ are equal, and they discard the mismatched bits of $B_a$ and $B_b$. Then the remaining bits of $K_a$ and $K_b$, denoted as $K_a'$ and $K_b'$ with $K_{a,b}=K_a'=K_b'$.
\end{enumerate}

We re-introduce the basic BB84 protocol in an abstract way with more technical details as Fig.\ref{BB84} illustrates.

Now, we assume a special measurement operation $Rand[q;B_a]$ which create a string of $n$ random bits $B_a$ from the $q$ quantum system, and the same as $Rand[q;K_a]$, $Rand[q';B_b]$. $M[q;K_b]$ denotes the Bob's measurement operation of $q$. The generation of $n$ qubits $q$ through two quantum operations $Set_{K_a}[q]$ and $H_{B_a}[q]$. Alice sends $q$ to Bob through the quantum channel $Q$ by quantum communicating action $send_{Q}(q)$ and Bob receives $q$ through $Q$ by quantum communicating action $receive_{Q}(q)$. Bob sends $B_b$ to Alice through the public channel $P$ by classical communicating action $send_{P}(B_b)$ and Alice receives $B_b$ through channel $P$ by classical communicating action $receive_{P}(B_b)$, and the same as $send_{P}(B_a)$ and $receive_{P}(B_a)$. Alice and Bob generate the private key $K_{a,b}$ by a classical comparison action $cmp(K_{a,b},K_a,K_b,B_a,B_b)$. Let Alice and Bob be a system $AB$ and let interactions between Alice and Bob be internal actions. $AB$ receives external input $D_i$ through channel $A$ by communicating action $receive_A(D_i)$ and sends results $D_o$ through channel $B$ by communicating action $send_B(D_o)$.

Then the state transition of Alice can be described by qACP as follows.

\begin{eqnarray}
&&A=\sum_{D_i\in \Delta_i}receive_A(D_i)\cdot A_1\nonumber\\
&&A_1=Rand[q;B_a]\cdot A_2\nonumber\\
&&A_2=Rand[q;K_a]\cdot A_3\nonumber\\
&&A_3=Set_{K_a}[q]\cdot A_4\nonumber\\
&&A_4=H_{B_a}[q]\cdot A_5\nonumber\\
&&A_5=send_Q(q)\cdot A_6\nonumber\\
&&A_6=receive_P(B_b)\cdot A_7\nonumber\\
&&A_7=send_P(B_a)\cdot A_8\nonumber\\
&&A_8=cmp(K_{a,b},K_a,K_b,B_a,B_b)\cdot A\nonumber
\end{eqnarray}

where $\Delta_i$ is the collection of the input data.

And the state transition of Bob can be described by qACP as follows.

\begin{eqnarray}
&&B=receive_Q(q)\cdot B_1\nonumber\\
&&B_1=Rand[q';B_b]\cdot B_2\nonumber\\
&&B_2=M[q;K_b]\cdot B_3\nonumber\\
&&B_3=send_P(B_b)\cdot B_4\nonumber\\
&&B_4=receive_P(B_a)\cdot B_5\nonumber\\
&&B_5=cmp(K_{a,b},K_a,K_b,B_a,B_b)\cdot B_6\nonumber\\
&&B_6=\sum_{D_o\in\Delta_o}send_B(D_o)\cdot B\nonumber
\end{eqnarray}

where $\Delta_o$ is the collection of the input data.

The send action and receive action of the same data through the same channel can communicate each other, otherwise, a deadlock $\delta$ will be caused. We define the following communication functions.

\begin{eqnarray}
&&\gamma(send_Q(q),receive_Q(q))\triangleq c_Q(q)\nonumber\\
&&\gamma(send_P(B_b),receive_P(B_b))\triangleq c_P(B_b)\nonumber\\
&&\gamma(send_P(B_a),receive_P(B_a))\triangleq c_P(B_a)\nonumber
\end{eqnarray}

Let $A$ and $B$ in parallel, then the system $AB$ can be represented by the following process term.

$$\tau_I(\partial_H(A\parallel B))$$

where $H=\{send_Q(q),receive_Q(q),send_P(S_b),receive_P(S_b),send_P(S_a),receive_P(S_a)\}$ and $I=\{Rand[q;B_a], Rand[q;K_a], Set_{K_a}[q], H_{B_a}[q], Rand[q';B_b], M[q;K_b], c_Q(q), c_P(B_b),\\ c_P(B_a), cmp(K_{a,b},K_a,K_b,B_a,B_b)\}$.

Then we get the following conclusion.

\textbf{Theorem \ref{Verification}.1}. The basic BB84 protocol $\tau_I(\partial_H(A\parallel B))$ exhibits desired external behaviors.

\begin{proof}
\begin{eqnarray}
&&\partial_H(A\parallel B)=\sum_{D_i\in \Delta_i}receive_A(D_i)\cdot\partial_H(A_1\parallel B)\nonumber\\
&&\partial_H(A_1\parallel B)=Rand[q;B_a]\cdot\partial_H(A_2\parallel B)\nonumber\\
&&\partial_H(A_2\parallel B)=Rand[q;K_a]\cdot\partial_H(A_3\parallel B)\nonumber\\
&&\partial_H(A_3\parallel B)=Set_{K_a}[q]\cdot\partial_H(A_4\parallel B)\nonumber\\
&&\partial_H(A_4\parallel B)=H_{B_a}[q]\cdot\partial_H(A_5\parallel B)\nonumber\\
&&\partial_H(A_5\parallel B)=c_Q(q)\cdot\partial_H(A_6\parallel B_1)\nonumber\\
&&\partial_H(A_6\parallel B_1)=Rand[q';B_b]\cdot\partial_H(A_6\parallel B_2)\nonumber\\
&&\partial_H(A_6\parallel B_2)=M[q;K_b]\cdot\partial_H(A_6\parallel B_3)\nonumber\\
&&\partial_H(A_6\parallel B_3)=c_P(B_b)\cdot\partial_H(A_7\parallel B_4)\nonumber\\
&&\partial_H(A_7\parallel B_4)=c_P(B_a)\cdot\partial_H(A_8\parallel B_5)\nonumber\\
&&\partial_H(A_8\parallel B_5)=cmp(K_{a,b},K_a,K_b,B_a,B_b)\cdot\partial_H(A\parallel B_5)\nonumber\\
&&\partial_H(A\parallel B_5)=cmp(K_{a,b},K_a,K_b,B_a,B_b)\cdot\partial_H(A\parallel B_6)\nonumber\\
&&\partial_H(A\parallel B_6)=\sum_{D_o\in\Delta_o}send_B(D_o)\cdot\partial_H(A\parallel B)\nonumber
\end{eqnarray}

Let $\partial_H(A\parallel B)=\langle X_1|E\rangle$, where E is the following guarded linear recursion specification:

\begin{eqnarray}
&&\{X_1=\sum_{D_i\in \Delta_i}receive_A(D_i)\cdot X_2,X_2=Rand[q;B_a]\cdot X_3,X_3=Rand[q;K_a]\cdot X_4,\nonumber\\
&&X_4=Set_{K_a}[q]\cdot X_5,X_5=H_{B_a}[q]\cdot X_6,X_6=c_Q(q)\cdot X_7),\nonumber\\
&&X_7=Rand[q';B_b]\cdot X_8,X_8=M[q;K_b]\cdot X_9,X_9=c_P(B_b)\cdot X_{10},X_{10}=c_P(B_a)\cdot X_{11},\nonumber\\
&&X_{11}=cmp(K_{a,b},K_a,K_b,B_a,B_b)\cdot X_{12},X_{12}=cmp(K_{a,b},K_a,K_b,B_a,B_b)\cdot X_{13},\nonumber\\
&&X_{13}=\sum_{D_o\in\Delta_o}send_B(D_o)\cdot X_1\}\nonumber
\end{eqnarray}

Then we apply abstraction operator $\tau_I$ into $\langle X_1|E\rangle$.

\begin{eqnarray}
\tau_I(\langle X_1|E\rangle)
&=&\sum_{D_i\in \Delta_i}receive_A(D_i)\cdot\tau_I(\langle X_2|E\rangle)\nonumber\\
&=&\sum_{D_i\in \Delta_i}receive_A(D_i)\cdot\tau_I(\langle X_3|E\rangle)\nonumber\\
&=&\sum_{D_i\in \Delta_i}receive_A(D_i)\cdot\tau_I(\langle X_4|E\rangle)\nonumber\\
&=&\sum_{D_i\in \Delta_i}receive_A(D_i)\cdot\tau_I(\langle X_5|E\rangle)\nonumber\\
&=&\sum_{D_i\in \Delta_i}receive_A(D_i)\cdot\tau_I(\langle X_6|E\rangle)\nonumber\\
&=&\sum_{D_i\in \Delta_i}receive_A(D_i)\cdot\tau_I(\langle X_7|E\rangle)\nonumber\\
&=&\sum_{D_i\in \Delta_i}receive_A(D_i)\cdot\tau_I(\langle X_8|E\rangle)\nonumber\\
&=&\sum_{D_i\in \Delta_i}receive_A(D_i)\cdot\tau_I(\langle X_9|E\rangle)\nonumber\\
&=&\sum_{D_i\in \Delta_i}receive_A(D_i)\cdot\tau_I(\langle X_{10}|E\rangle)\nonumber\\
&=&\sum_{D_i\in \Delta_i}receive_A(D_i)\cdot\tau_I(\langle X_{11}|E\rangle)\nonumber\\
&=&\sum_{D_i\in \Delta_i}receive_A(D_i)\cdot\tau_I(\langle X_{12}|E\rangle)\nonumber\\
&=&\sum_{D_i\in \Delta_i}receive_A(D_i)\cdot\tau_I(\langle X_{13}|E\rangle)\nonumber\\
&=&\sum_{D_i\in \Delta_i}\sum_{D_o\in\Delta_o}receive_A(D_i)\cdot send_B(D_o)\cdot \tau_I(\langle X_1|E\rangle)\nonumber
\end{eqnarray}

We get $\tau_I(\langle X_1|E\rangle)=\sum_{D_i\in \Delta_i}\sum_{D_o\in\Delta_o}receive_A(D_i)\cdot send_B(D_o)\cdot \tau_I(\langle X_1|E\rangle)$, that is, $\tau_I(\partial_H(A\parallel B))=\sum_{D_i\in \Delta_i}\sum_{D_o\in\Delta_o}receive_A(D_i)\cdot send_B(D_o)\cdot \tau_I(\partial_H(A\parallel B))$. So, the basic BB84 protocol $\tau_I(\partial_H(A\parallel B))$ exhibits desired external behaviors.
\end{proof}

\section{Extensions -- Renaming Operator}\label{Extensions}

One of the most Fascinating characteristics is the modularity of ACP, that is, ACP can be extended easily. Through out this paper, we can see that qACP also inherent the modularity characteristics of ACP. By introducing new operators or new constants, qACP can have more properties. It is already proved that ACP or qACP possibly has the same expressive power as a Turing machine \cite{Consistency}. Though extensions can not improve the expressive power of qACP, but they provide qACP an elegant fashion to express a new property.

In this section, we take an example of renaming operator which is used to rename the atomic quantum operations.

\subsection{Transition Rules of Renaming Operators}

Renaming operator $\rho_f(t)$ renames all actions in process term $t$, and the change of the quantum state is consistent, which is expressed by the following two transition rules.

$$\frac{\langle x,\varrho\rangle\xrightarrow{\upsilon}\langle\surd,\varrho'\rangle}{\langle \rho_f(x),\varrho\rangle\xrightarrow{f(\upsilon)}\langle\surd,\varrho'\rangle}$$

$$\frac{\langle x,\varrho\rangle\xrightarrow{\upsilon}\langle x',\varrho'\rangle}{\langle \rho_f(x),\varrho\rangle\xrightarrow{f(\upsilon)}\langle\rho_f(x'),\varrho'\rangle}$$

\textbf{Theorem \ref{Extensions}.1}. $\textrm{AQCP}_{\tau}$ with guarded linear recursion and renaming operators is a conservative extension of $\textrm{AQCP}_{\tau}$ with guarded linear recursion.

\begin{proof}
The corresponding TSS of $\textrm{AQCP}_{\tau}$ guarded linear recursion and renaming operators is a conservative extension of that of $\textrm{AQCP}_{\tau}$ with guarded linear recursion. That means that $\textrm{AQCP}_{\tau}$ guarded linear recursion and renaming operators is a conservative extension of $\textrm{AQCP}_{\tau}$ with guarded linear recursion.
\end{proof}

\textbf{Theorem \ref{Extensions}.2}. Quantum rooted branching bisimulation equivalence is a congruence with respect to $\textrm{AQCP}_{\tau}$ with guarded linear recursion and renaming operators.

\begin{proof}
The structural part of QTSSs for $\textrm{AQCP}_{\tau}$ guarded linear recursion and renaming operators are all in RBB cool format by incorporating the successful termination predicate $\downarrow$ in the transition rules, so rooted branching bisimulation equivalence that they induce is a congruence. According to the definition of quantum rooted branching bisimulation, quantum rooted branching bisimulation equivalence that QTSSs for $\textrm{AQCP}_{\tau}$ guarded linear recursion and renaming operators induce is also a congruence.
\end{proof}

\subsection{Axioms for Renaming Operators}

The axioms for renaming operator is shown in Table \ref{AxiomForR}.

\begin{center}
\begin{table}
  \begin{tabular}{@{}ll@{}}
\hline No. &Axiom\\
  QRN1 & $\rho_f(\upsilon)=f(\upsilon)$\\
  QRN2 & $\rho_f(\delta)=\delta$\\
  QRN3 & $\rho_f(x+y)=\rho_f(x)+\rho_f(y)$\\
  QRN4 & $\rho_f(x\cdot y)=\rho_f(x)\cdot\rho_f(y)$\\
\end{tabular}
\caption{Axioms for renaming}
\label{AxiomForR}
\end{table}
\end{center}

\textbf{Theorem \ref{Extensions}.3}. $\mathcal{E}_{\textrm{AQCP}_{\tau}}$ + RSP + RDP + CFAR + QRN1-QRN4 is sound for $\textrm{AQCP}_{\tau}$ with guarded linear recursion and renaming operators, modulo quantum rooted branching bisimulation equivalence.

\begin{proof}
Since quantum rooted branching bisimulation is both an equivalence and a congruence for $\textrm{AQCP}_{\tau}$ with guarded linear recursion and renaming operators, only the soundness of the first clause in the definition of the relation $=$ is needed to be checked. That is, if $s=t$ is an axiom in $\mathcal{E}_{\textrm{AQCP}_{\tau}}$ + RSP + RDP + CFAR + QRN1-QRN4 and $\sigma$ a closed substitution that maps the variable in $s$ and $t$ to basic quantum process terms, then we need to check that $\langle\sigma(s),\varrho\rangle\underline{\leftrightarrow}_{rb}\langle\sigma(t),\varsigma\rangle$.

Since axioms in $\mathcal{E}_{\textrm{AQCP}_{\tau}}$ + RSP + RDP + CFAR + QRN1-QRN4 (same as $\mathcal{E}_{\textrm{ACP}_{\tau}}$ + RSP + RDP + CFAR + QRN1-QRN4) are sound for $\textrm{ACP}_{\tau}$ with guarded linear recursion and renaming operators modulo rooted branching bisimulation equivalence, according to the definition of quantum rooted branching bisimulation, we only need to check if $\varrho'=\varsigma'$ when $\varrho=\varsigma$, where $\varrho$ evolves into $\varrho'$ after execution of $\sigma(s)$ and $\varsigma$ evolves into $\varsigma'$ after execution of $\sigma(t)$. We can find that every axiom in Table \ref{AxiomForR} meets the above condition.
\end{proof}

\textbf{Theorem \ref{Extensions}.4}. $\mathcal{E}_{\textrm{AQCP}_{\tau}}$ + RSP + RDP + CFAR + QRN1-QRN4 is complete for $\textrm{AQCP}_{\tau}$ with guarded linear recursion and renaming operators, modulo quantum rooted branching bisimulation equivalence.

\begin{proof}
To prove that $\mathcal{E}_{\textrm{AQCP}_{\tau}}$ + RSP + RDP + CFAR + QRN1-QRN4 is complete for $\textrm{AQCP}_{\tau}$ with guarded linear recursion and renaming operators modulo quantum rooted branching bisilumation equivalence, it means that $\langle s, \varrho\rangle\underline{\leftrightarrow}_{rb} \langle t,\varsigma\rangle$ implies $s=t$.

It was already proved that $\mathcal{E}_{\textrm{AQCP}_{\tau}}$ + RSP + RDP + CFAR + QRN1-QRN4 (same as $\mathcal{E}_{\textrm{ACP}_{\tau}}$ + RSP + RDP + CFAR + QRN1-QRN4) is complete for $\textrm{ACP}_{\tau}$ with guarded linear recursion and renaming operators modulo rooted branching bisimulation equivalence, that is, $s\underline{\leftrightarrow}_{rb} t$ implies $s=t$. $\langle s, \varrho\rangle\underline{\leftrightarrow}_{rb} \langle t,\varsigma\rangle$ with $\varrho=\varsigma$ means that $s\underline{\leftrightarrow}_{rb} t$ with $\varrho=\varsigma$ and $\varrho'=\varsigma'$, where $\varrho$ evolves into $\varrho'$ after execution of $s$ and $\varsigma$ evolves into $\varsigma'$ after execution of $t$, according to the definition of quantum rooted branching bisimulation equivalence. The completeness of $\mathcal{E}_{\textrm{AQCP}_{\tau}}$ + RSP + RDP + CFAR + QRN1-QRN4 for $\textrm{ACP}_{\tau}$ with guarded linear recursion and renaming operators modulo rooted branching bisimulation equivalence determines that $\mathcal{E}_{\textrm{AQCP}_{\tau}}$ + RSP + RDP + CFAR + QRN1-QRN4 is complete for $\textrm{AQCP}_{\tau}$ with guarded linear recursion and renaming operators modulo quantum rooted branching bisimulation equivalence.
\end{proof}

We can see that qACP with renaming operator and ACP with renaming operator can also be unified under the framework of quantum process configuration $\langle p, \varrho\rangle$.

\section{Conclusions}\label{Conclusions}

In this paper, we extend the traditional structural operational semantics under the framework of quantum process configuration $\langle p,\varrho\rangle$ to support quantum processes. Based on the relationship between quantum bisimularity and classical bisimularity, we establish a series of axiomatization for quantum processes called qACP. We also unify qACP and classical ACP under the framework of quantum process configuration $\langle p,\varrho\rangle$. It makes qACP can adapt to all quantum communication protocols.

Now, we point out some future directions. (1) Quantum entanglement makes the processing of the silent step $\tau$ somewhat strange. The nature of influence of quantum entanglement for computation, especially for parallelism and concurrency, should be considered carefully and deeply in future, because quantum entanglement is unique for quantum mechanics. (2) Other novel framework representing quantum processes should be proposed, not only the quantum process configuration $\langle p,\varrho\rangle$. New framework will unify quantum computing and classical computing in a new way, which maybe capture the nature of quantum computing more naturally. (3) qACP inherits the modularity of ACP and makes it can be extended in an elegant fashion, in future, more properties can be extended in qACP. (4) The axiomatization of qACP can be used to verify most quantum communication protocols easily and widely in future.

\label{lastpage}


\begin{thebibliography}{Lam94}
  \bibitem{KB} D.E. Knuth and P.B. Bendix.:
    \emph{Simple word problems in universal algebras.}
    Computational Problems in Abstract Algebra, Pergamon Press, 1970, 263--297.
    
  \bibitem{PA} J. C. M. Baeten.:\emph{A brief history of process algebra.}
    Theor Comput Sci In Process Algebra, 2005, 335(2-3): 131--146.

  \bibitem{CCS} R. Milner.: \emph{Communication and Concurrency.}
    Prentice Hall, 1989.
    
  \bibitem{CCS2} R. Milner and J. Parrow and D. Walker.: \emph{A calculus of mobile processes, Parts I and II.}
    Information and Computation, 1992, 100(1992): 1--77.
    
  \bibitem{CSP} C. A. R. Hoare.: \emph{Communicating sequential processes.}
    http://www.usingcsp.com/. 1985.
    
  \bibitem{ACP} W. Fokkink.: \emph{Introduction to process algebra 2nd ed.}
    Springer-Verlag, 2007.
    
  \bibitem{SOS} G. D. Plotkin.: \emph{A structural approach to operational semantics.}
    Aarhus University, Tech. Report DAIMIFN-19, 1981.
    
  \bibitem{Consistency} J. C. M. Baeten and J. A. Bergstra and J. W. Klop.: \emph{On the consistency of Koomen's fair abstraction rule.}
    Theoretical Computer Science, 1987, 51(1/2): 129--176.
    
  \bibitem{PSQP} Y. Feng and R. Y. Duan and Z. F. Ji and M. S. Ying.: \emph{Probabilistic bisimulations for quantum processes.}
    Information and Computation, 2007, 205(2007): 1608--1639.
    
  \bibitem{CQP} S. J. Gay and R. Nagarajan.: \emph{Communicating quantum processes.}
    Proceedings of the 32nd ACM Symposium on Principles of Programming Languages, Long Beach, California, USA, ACM Press, 2005: 145--157.
    
  \bibitem{CQP2} S. J. Gay and R. Nagarajan.: \emph{Typechecking communicating quantum processes.}
    Mathematical Structures in Computer Science, 2006, 16(2006): 375--406.
    
  \bibitem{QPA} P. Jorrand and M. Lalire.: \emph{Toward a quantum process algebra.}
    Proceedings of the 1st ACM Conference on Computing Frontiers, Ischia, Italy, ACM Press, 2005: 111--119.
    
  \bibitem{QPA2} P. Jorrand and M. Lalire.: \emph{From quantum physics to programming languages: a process algebraic approach.}
    Lecture Notes in Computer Science, 2005, 3566(2005): 1--16.
    
  \bibitem{BC} M. Lalire.: \emph{Relations among quantum processes: Bisimilarity and congruence.}
    Mathematical Structures in Computer Science, 2006, 16(2006): 407--428.
    
  \bibitem{PAOS} M. Lalire and P. Jorrand.: \emph{A process algebraic approach to concurrent and distributed quantum computation: operational semantics.}
    Proceedings of the 2nd International Workshop on Quantum Programming Languages, TUCS General Publications, 2004: 109--126.
    
  \bibitem{qCCS} M. Ying and Y. Feng and R. Duan and Z. Ji.: \emph{An algebra of quantum processes.}
    ACM Transactions on Computational Logic (TOCL), 2009, 10(3): 1--36.
    
  \bibitem{SB} M. Hennessy and H. Lin.: \emph{Symbolic bisimulations.}
    Theoretical Computer Science, 1995, 138(2): 353--389.
    
  \bibitem{BQP} Y. Feng and R. Duan and and M. Ying.: \emph{Bisimulations for quantum processes.}
    Proceedings of the 38th ACM Symposium on Principles of Programming Languages (POPL 11), ACM Press, 2011: 523--534.
    
  \bibitem{BB84} C. H. Bennett and G. Brassard.: \emph{Quantum cryptography: Public-key distribution and coin tossing.}
    Proceedings of the IEEE International Conference on Computer, Systems and Signal Processing, 1984, 175--179.
    
  \bibitem{OBQP} Yuxin Deng and Yuan Feng.: \emph{Open bisimulation for quantum processes.}
    Manuscript, http://arxiv.org/abs/1201.0416, 2012.
    
  \bibitem{SBQP} Yuan Feng and Yuxin Deng and Mingsheng Ying.: \emph{Symbolic bisimulation for quantum processes.}
    Manuscript, http://arxiv.org/pdf/1202.3484, 2012.
    
  \bibitem{QCQI} M.A Nielsen and I. L Chuang.: \emph{Quantum Computation and Quantum Information.}
    Cambridge University Press, 2000.
\end{thebibliography}
\end{document}